\documentclass[reprint,
 amsmath,
 amssymb,
 aps, 
 pre,
 sort&compress,
 numbers,
 superscriptaddress
]{revtex4-2}

\usepackage{graphicx}
\usepackage{dcolumn}
\usepackage{bm}
\usepackage{xcolor}
\usepackage{hyperref}
\usepackage[french]{babel} 
\usepackage{float}

\begin{document}

\title{Viologen-based supramolecular crystal gels: \\gelation kinetics and sensitivity to temperature }


\author{Julien Bauland}
\affiliation{ENS de Lyon, CNRS, Laboratoire de physique, UMR 5672, F-69342 Lyon, France}%
\author{Vivien Andrieux}
\affiliation{ENS de Lyon, CNRS, LCH, UMR 5182, 69342, Lyon cedex 07, France}%
\author{Frédéric Pignon}
\affiliation{Univ. Grenoble Alpes, CNRS, Grenoble INP, LRP, F-38000 Grenoble, France}%
\author{Denis Frath} 
\email{denis.frath@ens-lyon.fr}
\affiliation{ENS de Lyon, CNRS, LCH, UMR 5182, 69342, Lyon cedex 07, France}%
\author{Christophe Bucher}
\email{christophe.bucher@ens-lyon.fr}
\affiliation{ENS de Lyon, CNRS, LCH, UMR 5182, 69342, Lyon cedex 07, France}%
\author{Thomas Gibaud}%
\email{thomas.gibaud@ens-lyon.fr}
\affiliation{ENS de Lyon, CNRS, Laboratoire de physique, UMR 5672, F-69342 Lyon, France}%

\date{\today}

\begin{abstract}
Supramolecular crystal gels, a subset of molecular gels, form through self-assembly of low molecular weight gelators into interconnecting crystalline fibers, creating a three-dimensional soft solid network.
This study focuses on the formation and properties of viologen-based supramolecular crystalline gels. It aims to answer key questions about the tunability of network properties and the origin of these properties through in-depth analyses of the gelation kinetics triggered by thermal quenching.
Experimental investigations, including UV-Vis absorption spectroscopy, rheology, microscopy and scattering measurements, contribute to a comprehensive and self-consistent understanding of the system kinetics. We confirm that the viologen-based gelators crystallize by forming nanometer radius hollow tube that assemble into micro to millimetric spherulites. We then show that the crystallization follows the Avrami theory and is based on pre-existing nuclei. We also establish that the growth is interface controlled 
leading to the hollow tubes to branch into spherulites with fractal structures. Finally, we demonstrate that the gel properties can be tuned depending on the quenching temperature. Lowering the temperature results in the formation of denser and smaller spherulites. In contrast, the gels elasticity is not significantly affected by the quench temperature, leading us to hypothesize that the spherulites densification occurs at the expense of the connectivity between spherulite
\end{abstract}

\maketitle
\section{Introduction}

Supramolecular crystal gels represent a subset of molecular gels~\cite{gels2006}, characterized by the self-assembly of low molecular weight gelator into crystalline fibers that interconnect to create a three-dimensional soft solid network~\cite{Liu2001, Liu2002, lescanne2003,  Li2010, andrieux2023}. These gels find applications in many fields such as food~\cite{pernetti2007}, cosmetics, drug delivery~\cite{webber2017}, tissue engineering~\cite{saunders2019}, organic electronics~\cite{ajayaghosh2008} or chemical sensing~\cite{kartha2012}. They have also proved useful as templates for materials synthesis~\cite{van2003},  and for artwork conservation~\cite{carretti2010}. 
In addition, the reversible nature of the interactions involved in the assemblies gives these materials the ability to respond to different stimuli such as heat, light and sound.\cite{jones2016}

\begin{figure}
    \includegraphics[scale=0.52, clip=true, trim=0mm 0mm 0mm 0mm]{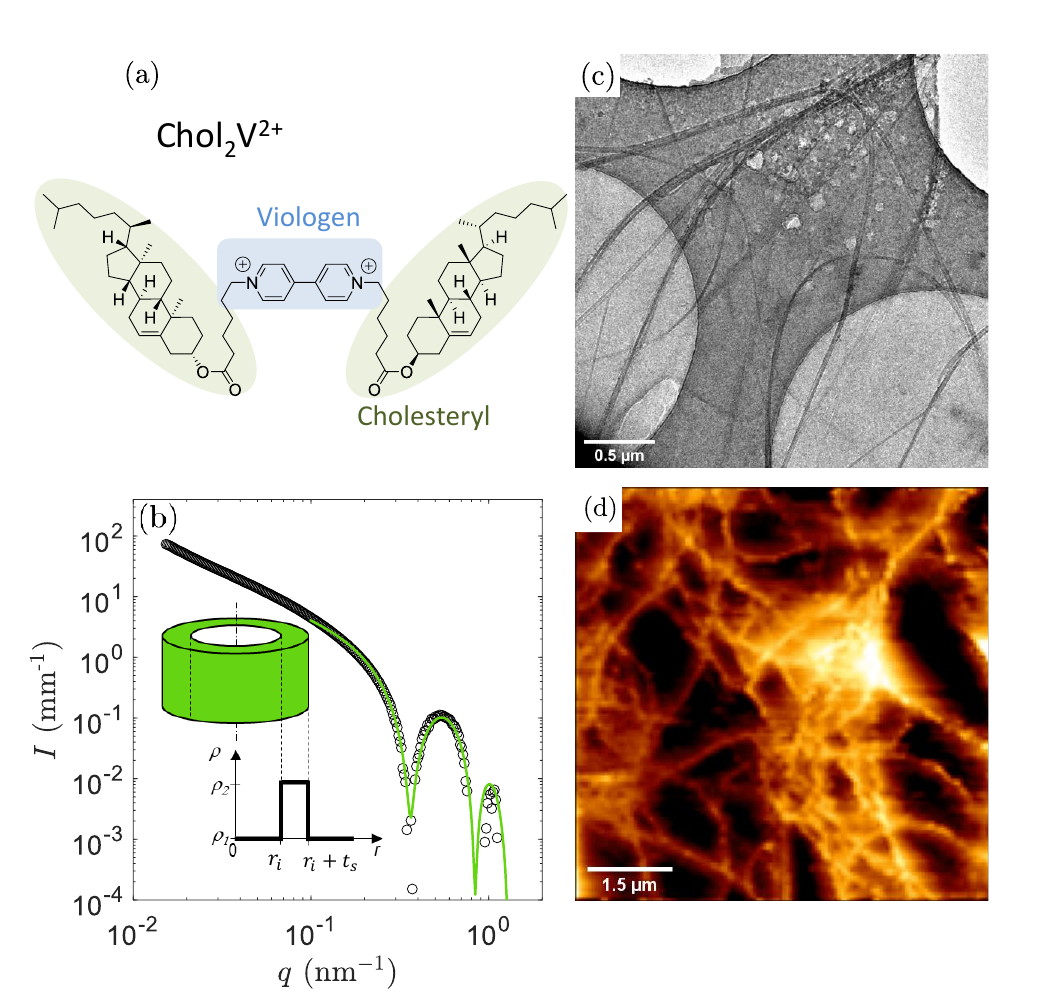}
    \centering
    \caption{ Characterization of the fibrous network in gel samples (a) Chemical structure of the viologen gelator $Chol_2V^{2+}$, composed of a 4,4’-bipyridinium unit bearing two cholesteryl substituants. (b) Microstructure of the gel assessed by small-angle X-ray scattering measurements. The scattering intensity $I(q)$ (circles) at high $q$ is best modeled by the form factor of a hollow cylinders (green curve).
    The inset figure represents the scattering length density profile of the hollow cylinder, with $\Delta \rho = \rho_1 - \rho_2 = 5.10^{-7}~\rm \mathring{A}^{-2}$ the difference of scattering length densities between the solvent and the viologen gelator. $r_i = 5.2~\rm nm$ and $t_s=2.6~\rm nm$ are the core radius and shell thickness of the cylinder, respectively.  (c)-(d) Fibrous structure of the gel observed by transmission electron cryomicroscopy and atomic force microscopy, respectively. 
    }
    \label{fig:fibers}
\end{figure}

We have recently demonstrated that supramolecular crystal gels exhibiting remarkable ionic conductivities  can be obtained from viologen-based low molecular weight gelator~\cite{andrieux2023}. 
As can be seen in Fig.~\ref{fig:fibers}, this gelator $Chol_2V^{2+}$ consists of two cholesteryl arms covalently linked to a 4,4'-bispyridinium plateform (best known as viologen. This dicationic molecule is not only redox- and photo-active, it is also capable of interacting with its environment through a wide range of “weak” van der Wall, electrostatics or Donor/Acceptor interactions. We have shown in previous works that a temperature quench triggers a spontaneous self-assembly of those monomers in hollow tubes.
Those tubes then grow to form spherulite ultimately percolating into a crystal gel network.

In general, viologen based gelators $V^{2+}$ present numerous advantages for creating conductive gels and hybrid materials. Beyond supplying two mobile counter anions, they have the potential to engage in electron transport through two consecutive one-electron reductions yielding stable radical cations $V^{+\bullet}$ and dianions $V^0$. These electron-responsive properties have already been extensively utilized in the development of molecular machines, electrochromic devices, and organic batteries~\cite{striepe2017,kathiresan2021}. Viologens have also been widely used as mediators in catalysis and sensing\cite{Fang2018,Dzhardimalieva2020} or more recently to enhance charge transport in single molecule~\cite{chen2021,li2021} and large-area molecular junctions~\cite{nguyen2018,han2020}. However, only a few examples of supramolecular gels involving viologen-based gelators can be found in the literature. Viologen-based gels have for instance been obtained upon exploiting their ability to form charge transfer complexes~\cite{Rao2010,Datta2017,Yuan2017,Yuan2016,Dhiman2020}. Gels have also been obtained by functionalizing the bipyridinium core with selected functions facilitating self-assembly, including van der Waals interactions~\cite{Xue2004}, H bonds~\cite{Suzuki2001}, Coulomb forces~\cite{Datta2016}, or metal-ligand bonds~\cite{Kahlfuss2018,Chowdhury2021,Roizard2022}.

In continuation of our previous works in the field~\cite{andrieux2023}, we now report  detailed analyses of the mechanical and structural properties of supramolecular crystal gels obtained by self-assembly of the cholesteryl-appended viologen $Chol_2V^{2+}$ (Fig.~\ref{fig:fibers}a). We have already demonstrated that the $Chol_2V^{2+}$-based solution undergoes crystallization into fibers, which grow into fractal spherulites, ultimately forming a gel. However, the crystallization mechanism and how the gel properties are influenced by temperature quench remain unknown. Can we tune the network properties, such as its elasticity and its structure, based on the temperature quench depth $T$? Can we rationalize the origin of these properties by studying the gelation kinetics, including the evolution of elastic and viscous moduli, the growth of crystalline spherulites, and the development of the network structure? This work aims to provide answers to these key questions and to improve our understanding of the complex phenomena involved in the formation of supramolecular gels.

As outlined in~\cite{yu2015}, the properties of supramolecular crystal gels depend mainly on three factors: (i) the way and type of crystals formed; (ii) the process by which the crystals grow and finally (iii) the way in which the crystals percolate and interact to establish a space-spanning elastic network.

The first stage is specific to the nature of the gelator molecule.
The early stage of crystallization is driven by thermodynamics~\cite{Liu2002}, specifically by the difference in chemical potential between the solution and the crystal: $\Delta \mu = \mu_{\text{sol}} - \mu_{\text{Xtal}}$. When $\Delta \mu > 0$, the system becomes supersaturated, a thermodynamic prerequisite for the onset and progression of a crystalline phase. In this metastable state, the concentration of the solute  $C$ exceeds the maximum solubility $C_{eq}$, i.e. the solution contains more dissolved solute than it would typically accommodate under equilibrium conditions at a specified temperature and pressure. The driving force for crystallization can then be expressed in terms of supersaturation, defined as $\sigma \simeq (C - C_{\text{eq}})/C_{\text{eq}}$. 
In the protocol we used, the solutions are prepared at $T=80$~°C to improve the solubility of the gelator, followed by cooling which reduces the solubility and causes crystallization and gelation.

A common approach for quantifying the early stage of crystallization kinetics involves use of the Avrami equation~\cite{avrami1941}. Introduced by the Soviet scientist M. Avrami, this equation has found widespread use in diverse fields, including materials science, metallurgy, chemistry, and physics~\cite{shirzad2023}. The equation establishes a relationship between the evolution of the crystal fraction $X$ and time $t$ through a reaction rate $k$, taking the form:
\begin{equation}
X=1-e^{-kt^n}
\label{eq:avrami}
\end{equation}
The Avrami exponent $n$~\cite{shirzad2023, gila2021} is a dimensionless parameter that encapsulates the nucleation growth mechanism through $n_G$, the dimensionality of growth through $d$, and the nucleation rate index through $n_N$:
\begin{multline}
  n = n_N + d n_G \\
  \text{with  }\begin{cases}
      n_G= 
     & \begin{cases} 0.5 & \text{diffusion controlled-growth}\\
      1 &\text{interface controlled-growth}
      \end{cases}
      \\
      d & \text{dimension of the growth}\\
      n_N= &  
      \begin{cases} 0 & \text{pre-existing nuclei}\\
      [0,1] & \text{decreasing nucleation rate}\\
      >1 & \text{increasing nucleation rate}
      \end{cases}
    \end{cases}
    \label{eq:n}
  \end{multline}
In our case, the viologen-based gelators crystallize as fibers~\cite{andrieux2023}. The nucleation rate index $n_N$ and whether fiber growth is interface-controlled ($n_G=1$) as in~\cite{ma2008,taguchi2001,heijna2007}, or diffusion-controlled ($n_G=0.5$) as in~\cite{mareau2005, zhu2007}, remain unknown.

In the second stage, fibers may grow to form more complex structures.
All our studies conducted so far on gels formed with $Chol_2V^{2+}$ support the conclusion that the crystal fibers undergo branching under the effect of crystallographic mismatches, ultimately forming spherulites~\cite{shi2009,Liu2002,wang2006,shtukenberg2012}.
This process, known as "Fiber Tip Branching", occurs at high supersaturation values, at which point nucleation and growth with crystallographic mismatches take place at the tips of the fibers, leading to branching. 
The resulting spherulites exhibit radial arms growing from a primary nucleation center branching into Cayley-like tree structures.

However, it remains unclear how these spherulites contribute to the formation of a 3D network. Do the branches entangle between neighboring spherulites, or do branches stick to one another due to attractive interactions?
Predictions~\cite{liu_sawant2002, lakshminarayanan2021, lam2010, nasr2021} were made 
at the macroscopic level to establish a direct link between the crystal gel structure and its mechanical properties. In this context, the spherulites are best described as Cayley trees of size $\xi$ and fractal dimension $d_f$ composed of branching segment of length $l$ and radius $a$. Based on simulations, Shi et al.~\cite{shi2009} assimilated the gel elasticity to that of the  spherulites. 
In~\cite{liu_sawant2002, lakshminarayanan2021, lam2010, nasr2021}, the authors proposed a model derived from the Avrami model~\cite{avrami1941} that links the  temporal evolution of the elastic modulus to the fractal dimension $d_f$ of the crystal gel network. The model establishes that $X$, the ratio of the volume fraction of crystal material at time $t$ to the volume fraction at $t \to \infty$, is linked to the evolution of the norm of the complex viscosity $\eta^*$ through the  equations $X = \frac{|\eta^*| - |\eta^*(0)|}{|\eta^*_{\infty}| - |\eta^*(0)|}$ and $-\ln(1-X) = k(t-t_g)^{d_f}$, assuming $n_N=0$ and $n_G=1$ et Eq.~\ref{eq:n}. 

\vspace{10pt}
In the present study, we utilize Small Angle X-ray Scattering (SAXS) and transmission electron cryomicroscopy (CryoTEM) to establish that the viologen-based gelators~\cite{andrieux2023} crystallize in the form of hollow tubes. Then, we establish the dependence of crystal supramolecular gel properties on the quench temperature: lowering the quench temperature prompts the growth of hollow tubes into smaller spherulites, a phenomenon observed through microscopy, and denser structures, as inferred from Ultra Small Angle Light Scattering (USALS). Interestingly, rheological measurements reveal seemingly constant gel elasticity across all tested temperature quenches.
Next, our focus shifts to gelation kinetics, aiming to precisely understand the origin of the influence of temperature quench on the final gel properties. We first focus on identifying the growth mechanism by tracking the spherulite growth under microscope. We then measure using UV-Vis absorption spectrosopy the crystal fraction $X$ and fit its time evolution with the Avrami equation. Those two complementary analysis allow us  to conclude that the crystallization is based on pre-existing nuclei, the growth is interface controlled and undergo branching due to crystallographic mismatch  to afford in fine spherutlites displaying fractal structures. Finally we explore the evolution of the viscous $G''$ and elastic $G'$  moduli. We demonstrate that the model derived from the Avrami equation is inadequate \cite{shi2009,Liu2002,wang2006,shtukenberg2012} for describing the time evolution of $G''$ and $G'$. Instead, the rheological properties must depend on both the elasticity of the spherulites and the nature of their connections.

\section{Material and methods}
\label{s:mm}
\subsection{Viologen-based gelator}
The viologen-based gelator used for all experiments is a cholesteryl-substituted viologen noted $Chol_2V^{2+}$ (Fig.~\ref{fig:fibers}a). It was obtained as a bromide salt $V(Br)_2$ using a protocol described previously ~\cite{andrieux2023}. The gelation ability of $Chol_2V^{2+}$ was first reported by Xue et al., who demonstrated that the gel fibrillar network could be efficiently used as a template for obtaining monodisperse mesoporous silica~\cite{Xue2004}. Our previous work showed that self-assembly of $Chol_2V^{2+}$ in 1-Pentanol relies on the formation of charge transfer (CT) complexes between $Chol_2V^{2+}$ (acceptor unit) and halide ions, $Br^-$ or $I^-$ (donor unit). 
The anions associated with the dicationic species $Chol_2V^{2+}$ play a major role in the self-assembly process and that the rheological, optical and electrical properties of the gels can be tuned by the addition of salts.
We also showed that the addition of lithium bis(trifluoromethanesulfonyl) ($LiTFSI$) helps to increase the solubility of $Chol_2V^{2+}$ in pentanol but that too much TFSI anions can prevent gelation. Addition of $TBAI$ also increases the amount of donors species available to form the $[Chol_2V^{2+}/I^-]$ CT complexes promoting gelation but yields opaque gels displaying high gelation temperature. 
An optimal composition of $TBAI$ and $LiTFSI$ has thus been determined to obtain a translucent and robust gel featuring a moderate gelation temperature and a high ionic conductivity~\cite{andrieux2023}. This study specifically focuses on this optimal composition consisting of $Chol_2V(Br)_2$ (1.3 mM, 0.2 wt\%), $TBAI$ (50 mM), and $LiTFSI$ (100 mM) in 1-Pentanol.
In all experiments, the sol-gel transition was induced by thermally quenching homogeneous solutions at a rate of $\sim 0.6^{\circ}$C/s from 80~$^{\circ}$C to a final temperature $T$ using a Peltier module. The shallower quench where carried out to a final temperature $T=50^{\circ}$C while the deeper quenches where carried out down to $T=20^{\circ}$C. The time $t=0$~s is defined as the moment when the final temperature is reached. 
\subsection{Transmission Electron Cryo-microscopy (CryoTEM)}
 The mixture (0.2 wt\%, 100 mM $TFSI$, 50 mM $TBAI$ in pentanol) was heated until full solubilization using a hair dryer. 4~$\mu$L of the liquid sample were collected using an Eppendorf pippette cone. For the freezing process, the liquid sample was then quickly injected on a TEM grid placed in the thermalized chamber of the plunge freezing device (Thermo Fisher Vitrobot) at 40$^{\circ}$C. After around 10 s on the grid, the excess liquid was absorbed using filter paper (1s blotting time) and the sample was vitrified in liquid ethane at a temperature of -170 $^{\circ}$C Finally, the frozen samples were transferred to the TEM cryogenic sample holder (Fischione model 2550), and was maintained at a temperature of -172 $^{\circ}$C during the TEM observations. Experiments were performed under an accelerating voltage of 120 kV with a JEOL 1400Flash TEM equipped with a Gatan RIO 16 Mpx camera at Centre Technologique des Microstructures (CT$\mu$, Université Claude Bernard Lyon 1, Villeurbanne, France).
\subsection{Rheology}
Rheological characterizations were conducted using a stress-controlled rheometer (Anton Paar MCR301) employing a smooth parallel-plate geometry (diameter of 35 mm) with a 0.5 mm gap. Samples were preheated in the liquid state and introduced into the rheometer, which was thermalized at 85$^{\circ}C$. To prevent evaporation of the organic solvent, the outer edge of the geometry was filled with a fluorinated oleophobic liquid (Fomblin \textregistered Y 25/6) \cite{Newbloom2012}. The sol-gel transition was induced by thermally quenching the sample using a Peltier module embedded in the bottom plate. 
The loss $G''$ and elastic $G'$ moduli were recorded over time at a frequency $f = 0.1$ Hz and strain amplitude $\gamma = 0.1$~\%. The gel viscoelastic spectrum was recorded from $f=10^1$ to $10^{-2}$~Hz at a strain amplitude $\gamma = 0.1$~\%.
\subsection{Bright field microscopy}
Microscopy measurements were performed using a Nikon Ti-eclipse inverted light microscope operating in bright-field mode. A Zyla camera from Andor, equipped with a 10× objective, was employed to capture images of the samples. A 455 nm led lamp (M455, Thorlabs) was used as a light source, which significantly improved the contrast of the images due to the large absorption band of the gel centered at 440 nm \cite{andrieux2023}. Samples, initially in the liquid state, were loaded into custom optical cells consisting of a microscope borosilicate glass slide and a cover slip separated by a 200 µm thick double-sided adhesive spacer to ensure airtightness (Gene Frame, Thermo Scientific). The sol-gel transition was induced by thermally quenching the sample using a dedicated Peltier cell (MHCS120i, Microptik).
Image analysis was conducted using MATLAB (version: 9.13.0.2320565) and the Image Processing Toolbox (version 11.6). Images were segmented by first removing large background structures using a bottom-hat filter. Second, the branched structure of the growing spherulites was "filled" using successive morphological erosion and dilatation steps, which afterwards improved their contour detection. Finally, thresholding was performed using an hysteresis method. The edge detection of the spherulites in the binary images was then performed using the matlab function regionprops, allowing to calculate their area $A$ as function of time and an equivalent sphere radius $r = \sqrt{A / \pi}$.
We define $t^{mciro}_N$ as the earliest time at which a spherulite can be consistently detected. $t^{mciro}_N$ is influenced by the tracking algorithm's properties: its spatial sensitivity around r=25 $\mu$m and its ability to track the growth of a single spherulite over at least 20 frames.

\subsection{Small Angle X-ray Scattering (SAXS)}
Small Angle X-ray Scattering was conducted at the SWING beamline (Synchrotron Soleil, Saint-Aubin, France)~\cite{thureau2021}. The samples were enclosed in 1~mm borosilicate capillaries (WJM-Glass Muller GmbH) and positioned in a thermally controlled holder (Linkam Scientific Instruments, Tadworth, UK). The beam energy was set to 12 KeV. The camera (EigerX4M with a detection area: 162.5 $\times$ 155.2 mm$^2$ and a pixel size : 75 $\times$ 75 $\mu$m$^2$) was set at a distance of 1.5m from the sample. To derive the scattering intensity $I(q)$ as a function of norm of the scattering wave vector $q$, the two-dimensional scattering profiles of the viologen suspension and the background solvent were subtracted. The resulting scattering intensity remained isotropic, and an azimuthal average was performed to obtain a one-dimensional $I(q)$.
\subsection{Ultra Small Angle Light Scattering (USALS)}
Ultra small angle light scattering (USALS) was performed on a laboratory-made setup as described in \cite{Piau1999}. Briefly, the sample is illuminated with a 2 mW laser beam (He-Ne, $\lambda = 632.8$ nm) and the forward scattered light is focused with a Fresnel lens on a monochrome digital camera (AV MAKO G-419B POE, Allied Vision). 2D scattering images were processed using a specifically developed software (Vimba Matlab), allowing to regroup and average scattering data with SAXSutilities software \cite{Sztucki2007}. The scattering intensity remained isotropic for all samples and an azimuthal average was performed to obtain a one-dimensional intensity $I(q)$, with $q$ the scattering vector given by $q = (4\pi / \lambda) \sin(\theta / 2)$, where $\lambda$ the wavelength of the light and $\theta$ the scattering angle. 
Gel samples were prepared between slide and a cover slip and heated to 85$^{\circ}$C before being quenched at a final temperature $T$ for 3h. The gel samples were then mounted on the USALS setup and $I(q)$ was measured at room temperature. After a 3-hour quench at $T$, we verified, using optical microscopy, that the gel structure remained unchanged when left at room temperature. 
\subsection{UV-Vis absorption spectroscopy}
Kinetic evolution of the absorption spectra was recorded  during the sol-gel transition using an MCS 601 UV-NIR spectrophotometer, operating in the $\lambda = 250$ to $1000$ nm range. As for microscopy imaging, samples were prepared between a slide and a cover slip and the temperature was controlled with a Peltier cell. Absoprtion spectra of samples prepared for the  measurment of supersaturation concentration (see Appendix \ref{appendix:ceq}) were recorded using the same spectrophotometer operating in the $\lambda = 200$ to $800$ nm range. 

\section{Results and discussion}

\subsection{Self assembly at the local scale}
We have recently demonstrated that the gel forms macroscopic spherulites composed of nano-fibers~\cite{andrieux2023}. As illustrated in Fig.~\ref{fig:fibers}b, the fine-scale structure of these fibers is well characterized by SAXS measurements. The scattered intensity $I(q)$ plotted against the modulus of the scattering vector $q$ is accurately described, for $q > 0.1$ nm$^{-1}$, by a hollow core–shell cylinder model~\cite{livsey1987} with a well defined inner radius $r_i = 5.2$ nm and shell thickness $t_s = 2.6$ nm, resulting in a total diameter of 15.6~nm. This fit is corroborated by electron microscopy images where single or bundles of fibers can be observed [Fig.~\ref{fig:fibers}(c)]. Besides, electron microscopy images indicates that fiber branching occurs through an self-epitaxial nucleation mechanisms rather than  a fiber tip branching mechanism~\cite{Li2010}. The fibers are further structured into a percolated and fractal-like network, as revealed by in situ imaging of the gel using fluid immersion atomic force microscopy [Fig.~\ref{fig:fibers}(d)]. 

\subsection{Effect of the thermal quench on the gel structure and rheology}
\begin{figure}
    \includegraphics[scale=0.55, clip=true, trim=0mm 0mm 0mm 0mm]{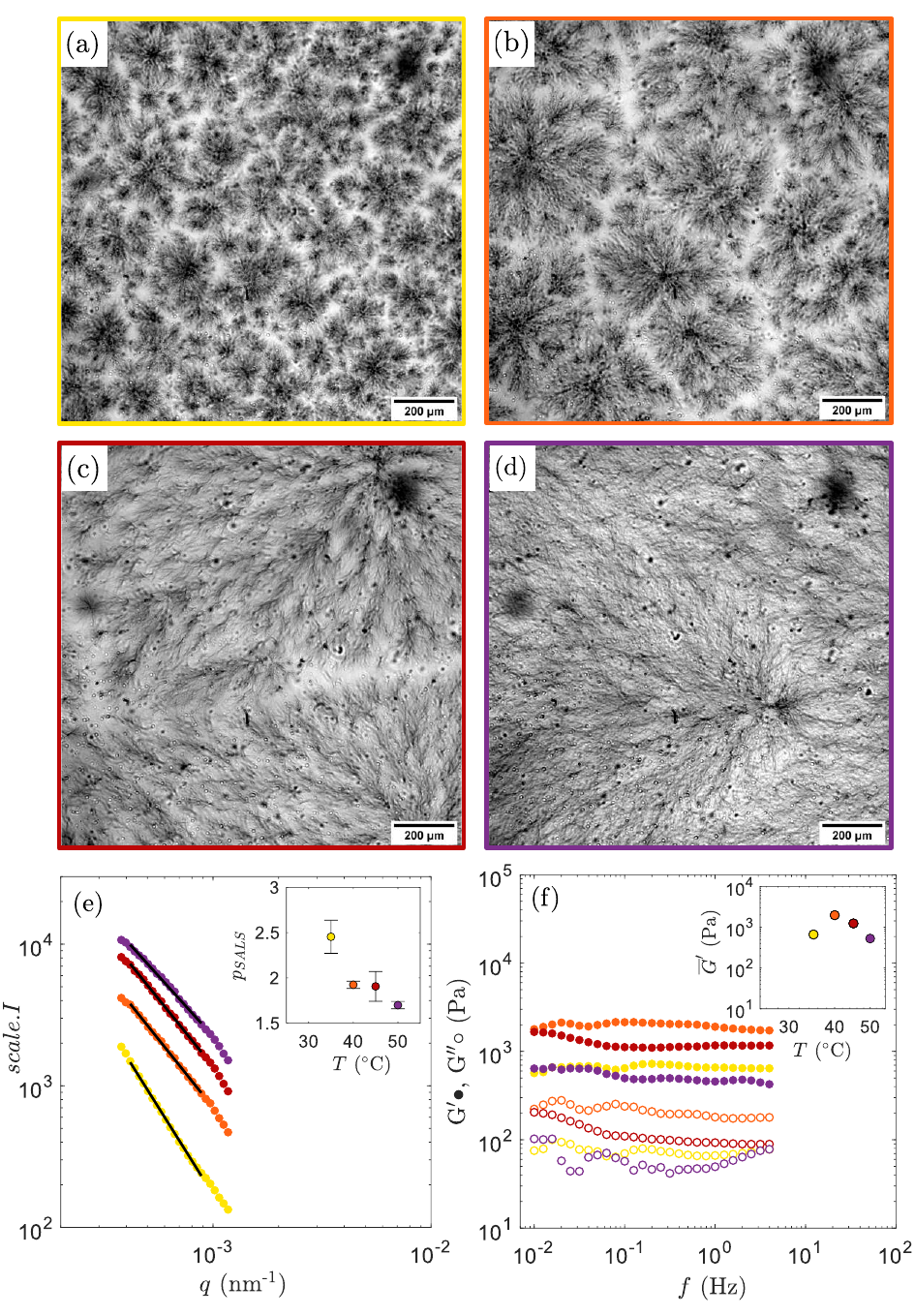}
    \centering
    \caption{(a)-(c) Microstructure of the gel determined from light microscopy after a thermal quench at T = 35 (yellow), 40 (orange) 45 (red) and 50$^{\circ}$C (violet), respectively. (e) Scattering curves $I(q)$ vs. $q$ of the gels obtained from light scattering. Black curve is the best power law fit $I(q) \sim q^{-p}$ for $4.10^{-4} < q < 1.10^{-3}$ nm$^{-1}$. Inset displays the the power exponent $p_s$ as function of the quench temperature. Errors bars are the standard deviation obtained from 10 replicates. (f) Storage (G$^{\prime}$) and loss (G$^{\prime\prime}$) moduli as function of the frequency ($f$). Inset displays the averaged storage modulus $\overline{G}^{\prime}(f)$ as function of the quench temperature. }
    \label{fig:quench}
\end{figure}
We first investigate the effect of the quench temperature $T$ on the final properties of the gels, i.e. after reaching a steady state as observed from microscopy imaging or rheometric measurements. The micrographs  displayed in Fig.~\ref{fig:quench}(a)-(d) reveal that the size of the spherulites decreases with $T$, down to few tens of microns at 35$^{\circ}$C. Besides, the small spherulites formed at low $T$ appear more contrasted and therefore denser. In Fig.~\ref{fig:quench}(e), we characterize the inner structure of the spherulites on length scales ranging from 2 to 500 \textmu m using USALS. The USALS scattering intensity $I(q)$ displayed against the scattering wave vector $q$ exhibits a power law behavior $I\propto q^{p_s}$ for  $4.10^{-4} < q < 1.10^{-3}$ nm$^{-1}$. The inset graph in Fig.~\ref{fig:quench}(e) shows that $p_s$ ranges between 1.7 and 2.5, indicating that the spherulites display a fractal organization with a fractal dimension $d_f = p_s$~\cite{Liu2001}. The later increases as $T$ decreases, indicating that the spherulites are more compact when formed at low $T$, which is consistent with the visual inspection of the spherulites in microscope images. The mechanical spectra displayed in Fig.~\ref{fig:quench}(f) show that for all quench temperatures, the gels display typical viscoelastic behavior of a soft solid, with the elastic modulus $G^{\prime}(f)$ showing no clear dependence with the frequency and dominating the viscous modulus $G^{\prime\prime}(f)$. Besides the elasticity of the gel does not show any clear dependence on $T$ and is about $10^3$ Pa. 

\subsection{Supersaturation as function of temperature}

\begin{figure}
    \includegraphics[scale=0.24, clip=true, trim=0mm 0mm 0mm 0mm]{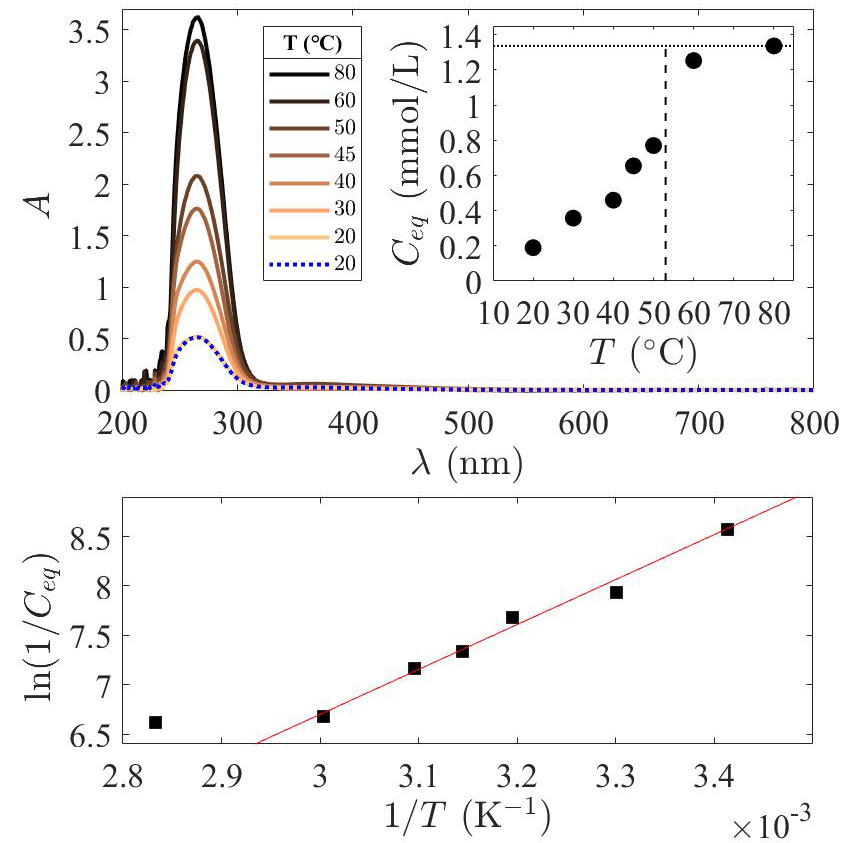}
    \centering
    \caption{Supersaturation measurements. (a) Corrected absorption spectra (see Appendix ~\ref{appendix:ceq}) recorded in a 1 mm optical cell (quartz) after applying the syringe (lines, temperature is color-coded from black 80~$^\circ$C to orange 20~$^\circ$C) and the centrifuge methods (dash blue at $T=20$~$^\circ$C) to gel samples quenched at different temperatures; The inset graph  shows the  maximum concentration $C_{eq}$ of the gelator deduced from the intensity of the absorption band at 265 nm using the absorption coefficient of $\epsilon_{\lambda=265nm}^{path=1 cm}=27000$~mol/L/cm. The dotted line is the initial concentration in gelators ($C=1.3$~mM) and the dashed line delimits the gel state obtained at low temperatures. (b) Evolution of $\ln(C_{eq})$ as a function of $1/T$. The red line is the best fit.}
    \label{fig:super}
\end{figure}

The changes observed in the structural and mechanical properties of the gel stem from the variation of the supersaturation $\sigma=(C-C_{eq})/C_{eq}$ with $T$, where the initial concentration of the gelator is $C=1.3$ mM. The maximum solubility at equilibrium $C_{eq}$ was measured using two different techniques (see Appendix~\ref{appendix:ceq}).
In the first technique, we centrifuged the mature gel at $10^4~g$ for 10 minutes at $T$, then recovered the supernatant at $C_{eq}$. In the second technique, we embedded the gel in a 1~mL syringe and recover the supernatant at $C_{eq}$ by pressing the gel through a filter (Appendix ~\ref{appendix:ceq}). Following mass conservation, we also deduced the concentration of $Chol_2V^{2+}$ integrated in the crystal network: $C_X=C-C_{eq}$.
In both cases, the concentration $C_{eq}$ was determined  by UV-vis absorption spectroscopy measurements based on the peak intensity at $\lambda=265$ nm. Fig.~\ref{fig:super}(a) demonstrates that both techniques yield similar results. In Fig.~\ref{fig:super}(a-Inset), we display the evolution of $C_{eq}$ as function of $T$. We observe, as expected, that $C_{eq}$ decreases when the quenches become deeper, i.e. as $T$ decreases. The Gel state is only observed below 50~$^{\circ}$C. 
In Fig. 3(b), we show that the solubility follows a Van’t Hoff law : $\ln(1/C_{eq}) = -\Delta_rH/RT + \Delta_r S/R$. According to the fit parameters, we found $\Delta_rH = -37\pm2$~kJ/mol and $\Delta_rS = -56 \pm 8$ J mol$^{-1}$K$^{-1}$. This thermodynamic data are consistent with thermodynamic data reported for the self-assembly of cholesterol derivatives ~\cite{murata1994,sarkar2020}.

\subsection{Gelation kinetics}
\label{sec:kinet}
In this section, we relate the gelation kinetics and the structure of the gels using crystal growth and nucleation models. Accordingly, we monitor the gelation at different quench temperature using microscopy, UV-Vis absorption spectroscopy and rheology measurements. This synergetic approach aims to identify the mechanisms at play during the gelation process. 
\begin{figure}[ht]
    \includegraphics[scale=0.45, clip=true, trim=0mm 0mm 0mm 0mm]{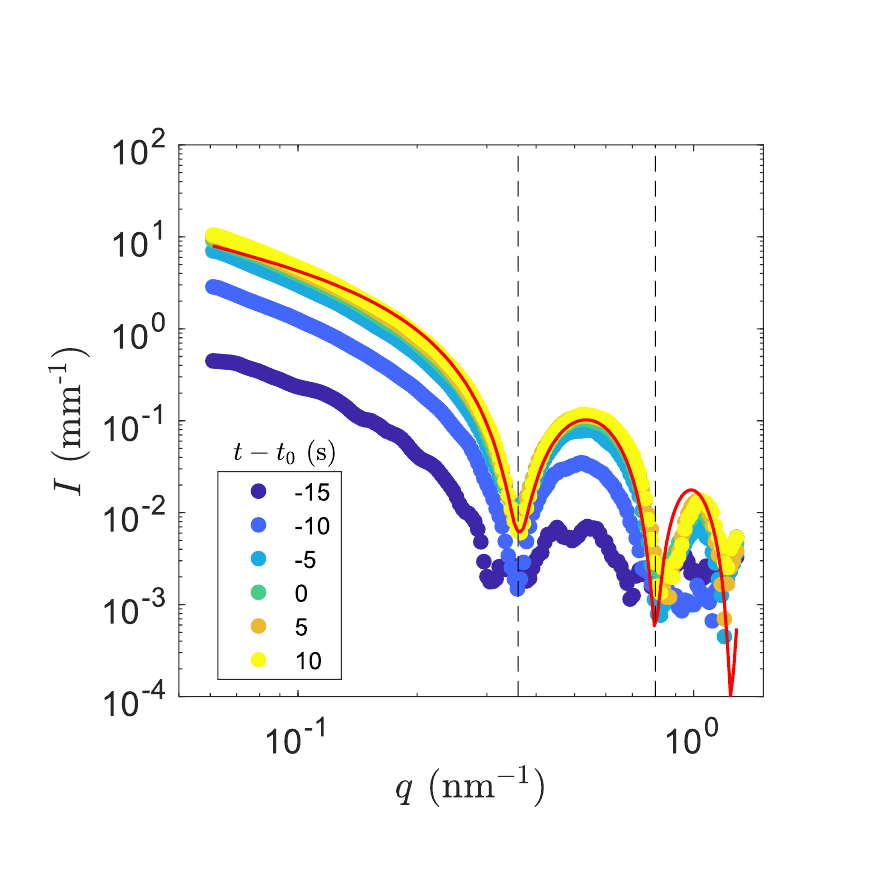}
    \centering
    \caption{Gelation kinetics assessed by SAXS upon cooling at 0.5$^{\circ}$C$/$s. The azimuthally averaged intensity $I(q)$ is displayed vs. the wave vector $q$ at different times during the cooling ramp to 20$^{\circ}$C. The time $t_0 = 140$ s corresponds to the time necessary to reach 20$^{\circ}$C from 90$^{\circ}$C. The first scattered signal was measured 15 s before reaching the quench temperature at T = 27$^{\circ}$C (bottom curve).
    The red curve represents the fit of the form factor for a core-shell cylinder with a core radius $r_i = 5.5~\rm$, a shell thickness $t_s = 2.9~\rm nm$ and a cylinder length $L > 700~\rm nm$ for $t-t_0=10$~s. This fit with same parameters remains valid at all time scales.
    }
    \label{figS:saxs}
\end{figure}

\vspace{1em}
\noindent \textbf{Hollow fibers nucleation --} Using time-resolved SAXS during the temperature quench, we monitor the formation of the hollow fibers, Fig~\ref{figS:saxs}. The analysis of the scattering intensity $I(q)$ indicates that the gelators assemble directly into hollow tubes. Their geometrical properties, the core radius $r_i = 5.5~\rm$  and the shell thickness $t_s = 2.9~\rm nm$ remains unchanged during the kinetics. This evidences that crystalline fibers self-assemble into hollow cylinders, without intermediary structure allowing us to rule out mechanisms where the gelators self-assembles into large 2D sheets that subsequently fold into tubes~\cite{lai2016,landman2018}. We also observe that the intensity of the scattering intensity increases with time indicating that the fibers concentration increases to rapidly form spherulite on larger length scales.

\vspace{1em}
\noindent \textbf{Growth of the spherulites --} Moving to the macroscopic scale, we now focus on the spherulite growth. Within the nucleation and growth framework, the process can be either interface-controlled or diffusion-controlled~\cite{shirzad2023, gila2021}. In the next, the growth mechanism is determined by measuring the temporal evolution of the spherulite size $r\propto t^{n_G}$ from microscopy data. 

Individual spherulite radius $r$ were determined by automatic image processing as described in the Material and Methods section. The very low contrast of the spherulites formed at 50$^{\circ}$C did not allow for satisfying edge detection by our tracking algorithm, and only the gels formed at $T$ = 35, 40 and 45  °C were investigated (see Appendix~\ref{appendix:radius}). Besides, spherulites could only be detected when their size exceeded 20~$\mu$m. Consequently, we define the nucleation time $t^{micro}_N$ as the time needed to obtain spherulites with diameters of 50~\textmu m, well above the detection limit of the algorithm.

\begin{figure}
    \includegraphics[scale=0.42, clip=true, trim=5mm 0mm 0mm 0mm]{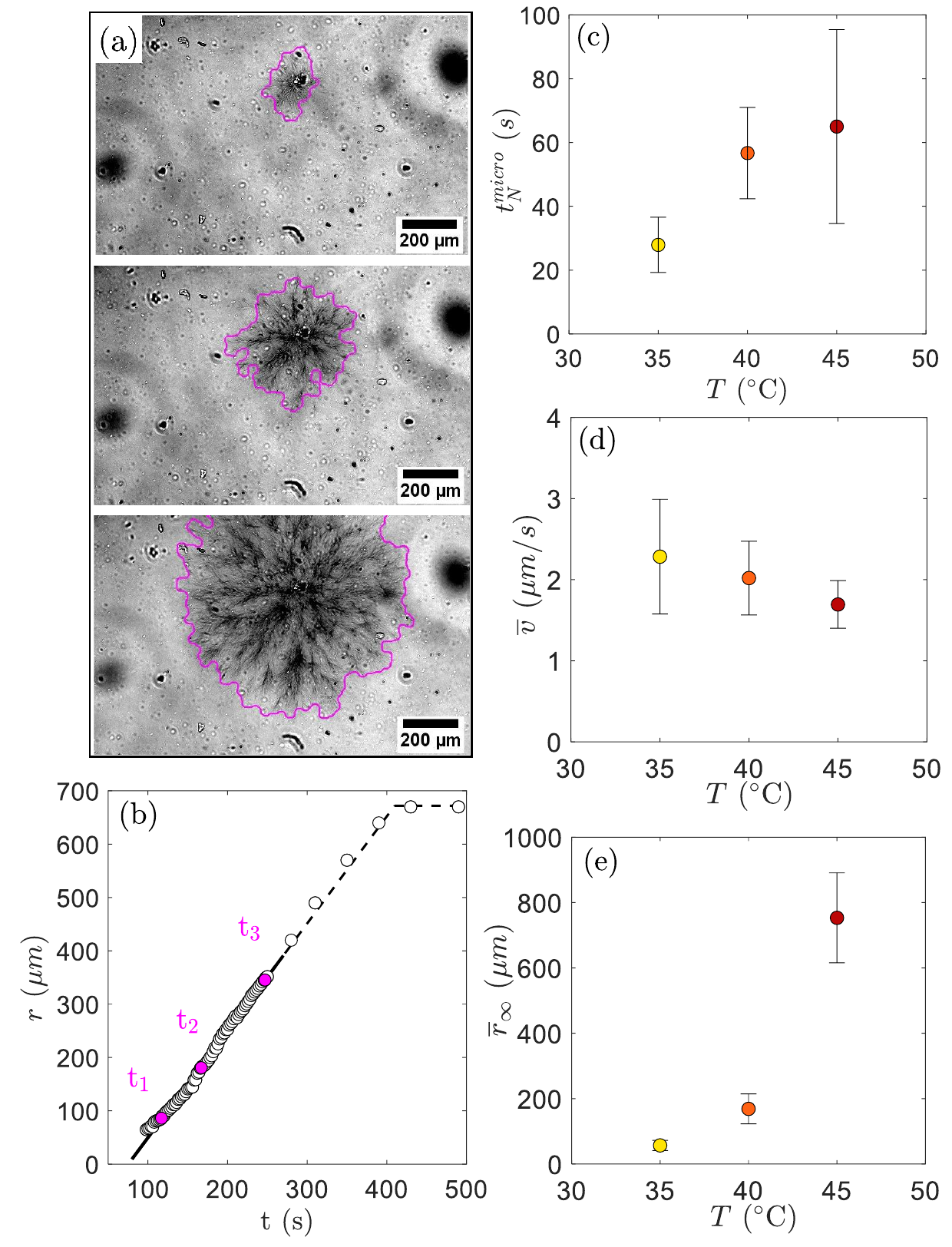}
    \centering
    \caption{Growth kinetics of spherulites assessed by bright field microscopy measurements. (a) Example of edge detection during the growth of a spherulite at $T = 45^{\circ}$C. From top to bottom, $t_1 < t_2 < t_3$. (b) Equivalent radius $r$ of the spherulite displayed in (a) as function of time. The times corresponding to each image of (a) are indicated in magenta. Plain black curve is the best linear fit of the data obtained from the tracking algorithms. Dotted black curves is an extrapolation of the linear growth of the spherulite up to its final size. (c) Nucleation time $t^{mciro}_N$ as function of the temperature of quench. (d) Growth rate $\overline{v}$ of individual spherulites as function of the temperature of quench. (e) Final radius $\overline{r}_{\infty}$ of the spherulites as function of the temperature of quench. Error bars are standard deviations.}
    \label{fig:microscopy}
\end{figure}

A typical evolution of a spherulite radius $r$ with time $t$ is shown in Fig.~\ref{fig:microscopy}(a)-(b) . We observe that $r$ increases linearly over time with a velocity $v$ [Fig.~\ref{fig:microscopy}(b)]. On larger timescales, the spherulite size saturates to reach a maximum value $\overline{r}_{\infty}$. The linear evolution of the spherulite size with time is indicative of an interface-controlled growth mechanisms with $n_G=1$, where the limiting step of the growth process is the assembly of viologen-gelators into crystalline fibers. In Fig.~\ref{fig:microscopy}(d), the average growth speed $\overline{v}$ is about 2 \textmu m.s$^{-1}$ and this value decreased slightly when the quench temperature increases [Fig.~\ref{fig:microscopy}(d)], highlighting that the diffusion of gelators is not the limiting step of the spherulite growth process. In contrast with $\overline{v}$, the nucleation time $t^{micro}_N$ and the average final spherulite radius $\overline{r}_{\infty}$ show a clear dependence with $T$, and increase when the quench temperature increases [Fig.~\ref{fig:microscopy}(c) and Fig.~\ref{fig:microscopy}(e)]. 

Those variations were also observed in other systems~\cite{ma2008,taguchi2001,heijna2007} and can be rationalized with variations of the driving force $\sigma$ upon changing the quench temperature. When the growth process is controlled by surface kinetics, it was concluded that variations in the driving force have much more effects on the nucleation frequency $J$ [s$^{-1}$] than on the growth speed~\cite{heijna2007}. This is consistent with the small effect of $T$ on $\overline{v}$ and the increase of $t^{micro}_N$ with $T$. Associated with the lower nucleation frequency at high $T$, the spherulites have less neighbors and are allowed to grow further, explaining the dependence of $\overline{r_{\infty}}$ with $T$. Besides, SALS experiments have shown that the fractal dimension of the spherulites decreases with $T$. For an heterogeneous nucleation mechanisms, the average branching length $l$ depends on the speed growth and on the nucleation frequency according $l = \overline{v} / J$. Consequently, as $\overline{v}$ is  roughly constant and $J$ decreases with $T$, the spherulite formed at high quench temperature are less branched with a lower fractal dimension. 

In the following sections, we relate the fractal dimension of the spherulites with the macroscopic kinetics of gelation assessed by UV-Vis absorption spectroscopy and rheology measurements.

\begin{figure*}
    \includegraphics[scale=0.50, clip=true, trim=0mm 0mm 0mm 0mm]{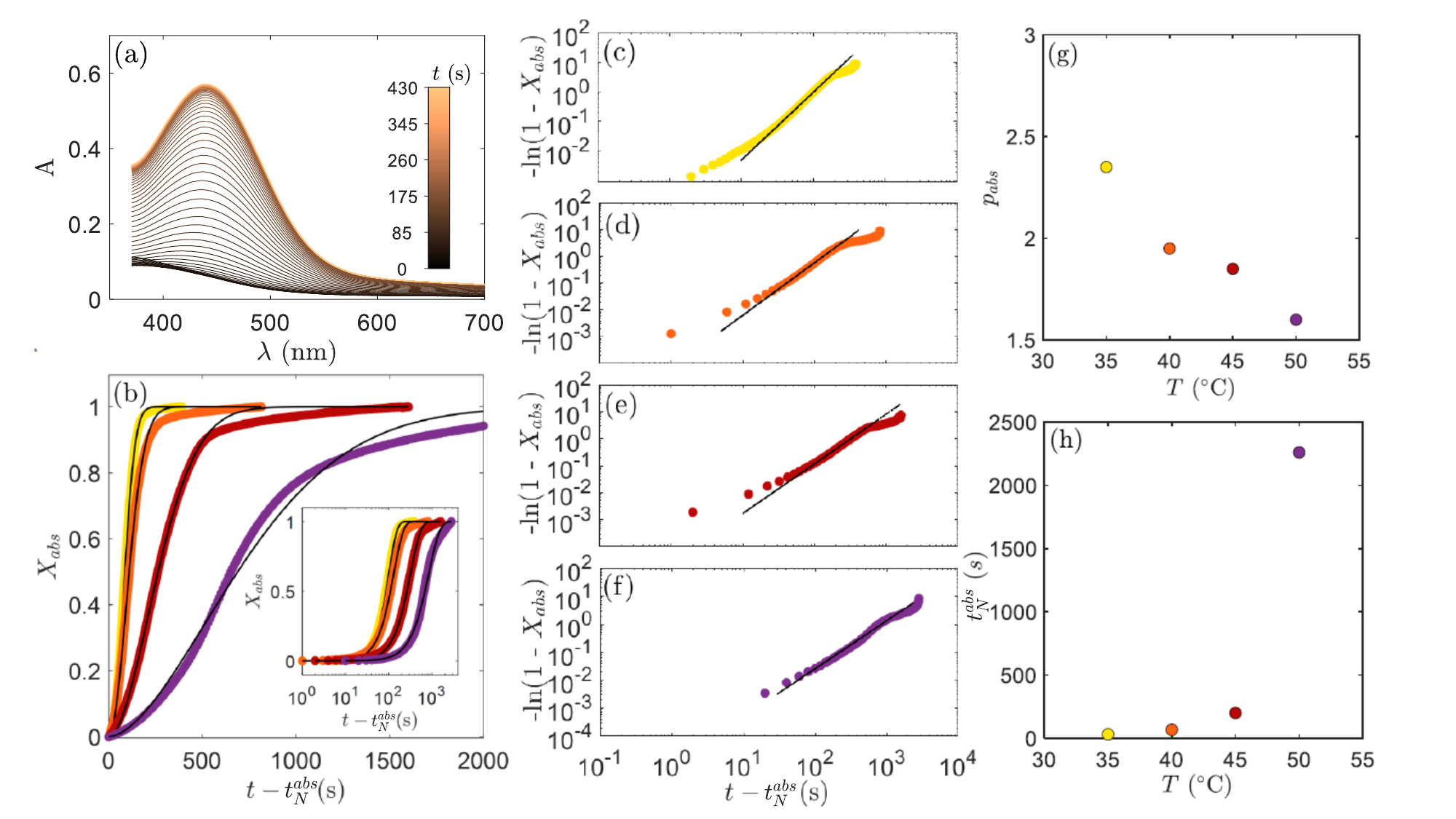}
    \centering
    \caption{Growth kinetics assessed by UV-vis absorption spectroscopy measurements. (a) Absorbance spectra recorded over time after thermal quench at 35$^{\circ}$C. The color codes for the time $t$. $t=0$~s correspond  to the time at which $T=35^{\circ}$C. (b) Crystallinity of the system $X_{abs}$ as a function of the time elapsed since the nucleation time $t_n$. Colors code for the temperature of quench: T = 35 (yellow), 40 (orange) 45 (red) and 50 $^{\circ}$C (violet). $X_{abs}$ is calculated from the time-evolution of the absorbance of the charge transfer band at $\lambda = 440$ nm, according to $X_{abs}(t)=[A_{CT}(t)-A_{CT}(t=0)]/[A_{CT}(t=\infty)-A_{CT}(t=0)]$, with $A_{CT}(t=0)$ and $A_{CT}(t=\infty)$ the absorbance at $t = t_N^{abs}$ and $t = \infty$. Black curves represent the Avrami relation which parameters are determined in (c)-(f). Inset displays a semi-log plot of the same data. (c)-(f) Avrami plots of the data displayed in (b). Black line is the best power law fit of the data. The power exponent $p_{abs}$ is displayed in (g) as function of the quench temperature $T$. (h) Nucleation time $t_N^{abs}$ as function of the temperature of quench.}
    \label{fig:UvKinetics}
\end{figure*}

\vspace{1em}
\noindent \textbf{Evolution of the crystal fraction --} In a crystallization process, the kinetics of the reaction can be related to the mechanisms of gelation, especially the dimension of growth, using the Avrami equation~\cite{avrami1941}. The extent of the reaction is typically assessed by the fraction of the crystalline phase with respect to the gelator concentration in solution.
In Fig.~\ref{fig:UvKinetics}(a), we display the evolution of the absorbance $A$ during the self-assembly of $Chol_2V^{2+}$ at 35$^{\circ}$C. We observe 3 main contributions to the absorption signal. In the low wavelength range ($\lambda<350$~nm) we observe the tail of the viologen-centered $\pi-\pi^*$ absorption band ($\lambda<270$~nm). At $\lambda=370$~nm and on short time scales, the large absorption band is attributed to solvated charge transfer species $[Chol_2V^{2+}/I^-]_{sol}$ \cite{monk1999}. Upon self-assembly of the gelator, this CT band grows and is progressively shifted to longer wavelength, resulting in a broad absorption peak centered at $\lambda = 440$~nm. The observed shift is attributed to the aggregation of charge transfer complexes $[Chol_2V^{2+}/I^-]_{agg}$ which are structuring elements of the supramolecular fibers. As previously described  by Nishikiori et. al. \cite{yoshikawa2005},  this red shift of the CT band can be attributed to an increase in the electronic delocalization upon aggregation of the $[Chol_2V^{2+}/I^-]$ CT complexes. In the following, we focus on the broad adsorption signal at $\lambda = 440$~nm.next, we focus on the broad adsorption peak at $\lambda = 440$~nm. We established that the absorbance of the charge transfer band at $\lambda = 440$~nm, $A_{CT}$, is proportional to the concentration of crystal fibers and follows the Beer-Lambert law, (see Appendix~\ref{appendix:Beer}).
The intensity of the UV absorption band at $\lambda = 440$~nm thus provides a direct estimation of the crystal fraction $X_{abs}(t)=\frac{A_{CT}(t)-A_{CT}(t=0)}{A_{CT}(t=\infty)-A_{CT}(t=0)}$.
We note $t_N^{abs}$, the nucleation time measured by UV absorption spectroscopy, define as $A_{CT}(t_N^{abs})= 0.05$ [Fig.~\ref{fig:UvKinetics}(h)]. The evolution of $X_{abs}$ as a function of the reduced time $t-t_N^{abs}$ is displayed in Fig.~\ref{fig:UvKinetics}(b) for different quench temperatures. 

Inspired by earlier reports~\cite{Liu2001,lakshminarayanan2021}, we use the Avrami representation $-\ln(1-X_{abs})$ vs. $t-t^{abs}_N$~[Fig.~\ref{fig:UvKinetics}(c)-(f)]. In this representation, we observe three regimes. At short time scales, $-\ln(1-X_{abs})$ evolves slowly with time. At intermediate time scales $-\ln(1-X_{abs})$ follows a power law evolution with an exponent $p_{abs}$: $\ln(1-X) = k(t-t^{abs}_N)^{p_{abs}}$. At long time scales $-\ln(1-X_{abs})$ tends to plateau as the gelation reaction is finished. 
The early time scales regime pertains to the initial stage of the self-assembly of the gelator into hollow tubes. We propose that a cooperative supramolecular polymerization mechanism occurs during this process, where the formation of a tubular structure strongly promotes the self-assembly of $Chol_2V^{2+}$\cite{de2009}, resulting in an apparent 'instantaneous growth' of the fibers. This hypotheses is backed up by two observations. First, when we plot $A(t)-A(t=0)$ (see Appendix~\ref{appendix:abs}), thus removing the contribution of the peak at $\lambda=370$~nm, we observe that both the shape and position along the $\lambda$-axis of the broad adsorption peak at $\lambda = 440$~nm remain unchanged. This suggests that at the interaction scales, the viologen gelators consistently interact in the same manner and likely do not assemble into structures other than hollow tubes in agreement with the time resolved SAXS presented in Fig.~\ref{figS:saxs}. 
The intermediate time scale regime allows us to determine the Avrami exponent $n$ in Eq.~\ref{eq:n}, here denoted $p_{abs}$. In Fig.~\ref{fig:UvKinetics}(g), $p_{asb}$ is consistent with the fractal dimension $d=p_{SALS}$ determined from scattering measurements, and decreases with $T$ [Fig.~\ref{fig:quench}(e)].  Given that $n_G = 1$ as determined from microscopy measurements, and since we identified that $d = p_{abs} = p_{SALS}$, this suggests that $n_N \approx 0$, indicating either pre-existing nuclei conditions formed during the early time scales or a decrease in the nucleation frequency. 

At this stage, we have validated the Avrami model and demonstrated that the viologen gelator crystallizes into hollow tubes, forming fractal spherulites. The growth conditions are determined by three factors: interface-controlled growth, the presence of pre-existing nuclei, and the relevant growth dimension being the fractal dimension of the spherulite.

\begin{figure*}
    \includegraphics[scale=0.55, clip=true, trim=0mm 0mm 0mm 0mm]{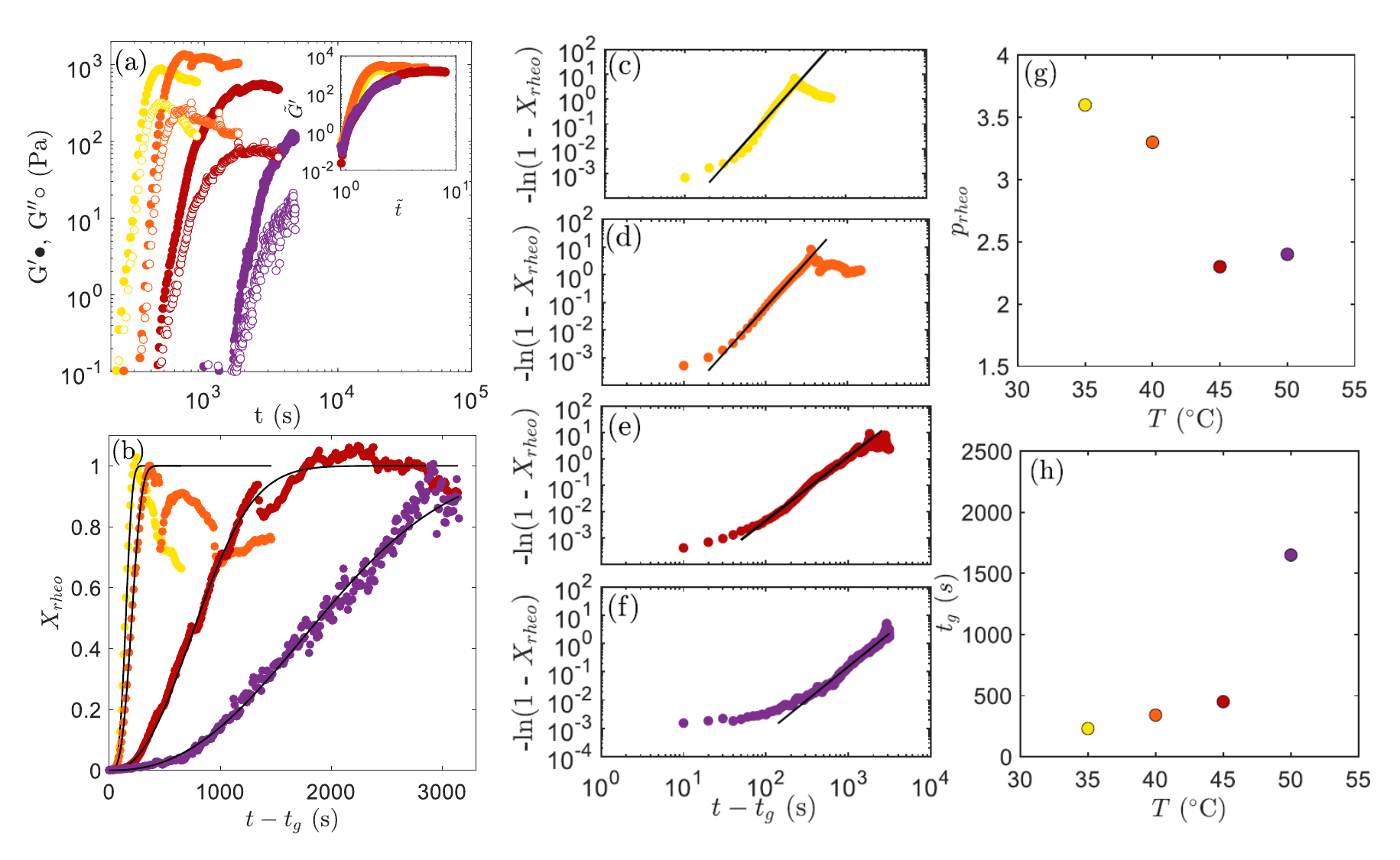}
    \centering
    \caption{Growth kinetics assessed by small-angle oscillatory rheometry measurements. (a) Time-evolution of the storage G$^{\prime}$ and loss G$^{\prime\prime}$ moduli. $t=0$~s corresponds to the time at which temperature is equal to the quench temperature. Colors code for the temperature of quench: T = 35 (yellow), 40 (orange) 45 (red) and 50$^{\circ}$C (violet). Inset: evolution of the rescaled elasticity $\tilde{G^{\prime}} = G^{\prime} / A_{CT}(t=\infty)$ with the rescaled time $t / t_n$. $\tilde{G^{\prime}}$ accounts for the variation crystallized gelator concentration with the temperature estimated by $A_{\infty}$, the final absorbance at $\lambda = 440$ nm. (b) Crystallinity of the system $X_{rheo}$ as function of the time elapsed since the gel time $t_g$, estimated by the cross-over between G$^{\prime}(t)$ and G$^{\prime\prime}(t)$. $X_{rheo}$ is calculated from the time-evolution of the complex viscosity $\eta^*$, according to $X_{rheo}(t) = (\eta^*(t) - \eta_0) / (\eta^*_{\infty} - \eta_0)$, with $eta_0$ the penthanol viscosity at the temperature of quench. Black curves represent the Avrami relation which parameters are determined in (c)-(f). (c)-(f) Avrami plots of the data displayed in (b). Black line is the best power law fit of the data. The power exponent $p_{rheo}$ is displayed in (g) as function of the temperature of quench. (h) Gel time $t_g$ as function of the temperature of quench.}
    \label{fig:rheoKinetics}
\end{figure*}

\vspace{1em}
\noindent \textbf{Evolution of the gel mechanical properties --} Finally, we assess the macroscopic kinetics of gelation using rheometric measurements. Fig.~\ref{fig:rheoKinetics}(a) displays the temporal evolution of the elastic modulus $G^{\prime}$ and the viscous modulus $G^{\prime\prime}$ for different quench temperatures. At short time scales, $G^{\prime\prime}$ dominates over $G^{\prime}$, while at longer time scales, $G^{\prime}$ surpasses $G^{\prime\prime}$. This behavior is characteristic of a sol-gel transition, and we estimate the gel time $t_g$ as the time for which $G^{\prime} > G^{\prime\prime}$. The gelation times displayed in Fig.~\ref{fig:rheoKinetics}(h) are of the same order of magnitude than the nucleation times determined from microscopy and UV-Vis absorption spectroscopy measurements, as expected from the large growth speed of the fibers ($\approx 2$ \textmu m.s$^{-1}$). At low quench temperatures, the viscoelastic moduli undergoes a small decrease after reaching a maximum value [Fig.~\ref{fig:rheoKinetics}(a)]. This behavior is attributed to negative normal forces, resulting from a contraction of the gel as reported for bio-polymer gels~\cite{mao2016}. In the inset graph of Fig.~\ref{fig:rheoKinetics}(a), we rescale the elastic modulus by the absorbance value $A_{CT}$ to account for variations in the concentration of crystallized gelator with $T$. When displayed against the rescaled time $t / t_g$, $\tilde{G^{\prime}}$ reaches a plateau value equals to about 1000 Pa for all quench temperatures. In previous reports, the elasticity of multi-domain networks increased with a decrease of $\overline{r}_{\infty}$~\cite{Li2010}. Given the strong variations of the final spherulite radius $\overline{r}_{\infty}$ and fractal dimension with $T$ [Fig.~\ref{fig:microscopy}(e) and Fig.~\ref{fig:quench}(e)], the present results suggest that the mesoscopic structure of the gel has little effect of its mechanical properties. However, potential differences may be masked by the small value of the gap ($e=500$ \textmu m). The final spherulite radius ($200 \leq \overline{r}_{\infty} \leq 800$ \textmu m) approximately matches the value of $e$, indicating that rheometric measurements predominantly assess the elasticity of the fibers or more localized interactions, such as fiber side branching, rather than the elasticity of the spherulites themselves~\cite{yu2015,Li2010,shi2009,yuan2011}.
In line with the work of Liu and coworkers~\cite{Liu2001,Liu2002,Liu2014}, the fraction of crystalline phase $X$ of a fiber network could be deduced from the norm of the complex viscosity $\eta^*=(G^{\prime}+iG^{\prime\prime})/\omega$, where $\omega$ is the angular frequency of small amplitude oscillation used to measure the viscoelastic moduli. 
By applying Einstein's relation ($\eta/\eta_0=(1+2.5\phi)$), one may connect the normalized viscosity and volume fraction of spherulite $\phi$.
Accordingly, the fraction of crystalline phase at a given time is calculated as $X_{rheo}= \frac{|\eta^*(t)| - |\eta^*(t=0)|}{|\eta^*(t=\infty)| - |\eta^*(t=0)|}$. In Fig.~\ref{fig:rheoKinetics}(b),  $X_{rheo}$ is displayed as a function of the reduced time $t-t_g$ for different quench temperatures. As with UV-vis absorption spectroscopy measurements, we use Avrami plots $-\ln(1-X_{rheo})$ vs $t-t_g$ to determine the Avrami exponents [Fig~\ref{fig:rheoKinetics}(c)-(e)]. In Fig.~\ref{fig:rheoKinetics}(g), we found that the exponents determined from rheological measurement $p_{rheo}$ are comprised between 3.6 and 2.3, and are thus overestimated in comparison with exponents determined from scattering and UV-Vis absorption spectroscopy measurements (Fig.~\ref{fig:quench}(d-inset) and Fig.~\ref{fig:UvKinetics}(g)). We cannot identify $p_{rheo}$ to $d$ as we identified $p_{abs}$ and $p_{SALS}$ to $d$ in the Avrami equation Eq.~\ref{eq:n}. We believe that these discrepancies stem 
from two assumptions. First, $t_g$ might not be the right characteristic time to normalise the $x$-axis in the Avrami representation. Replacing $t_g$ by $t_{N}^{abs}$ does not improve the results; in fact, it makes them even worse. Since $t_g$ is mostly larger than $t_{N}^{abs}$, plotting $-ln(1-X_{rheo})$  as a function of $t-t_{N}^{abs}$ leads to even higher values of $p_{rheo}$, which is already too large. Second, the assumption
of the linear relation between the viscosity of the solution and the volume fraction $\phi$, is not valid in the present system. Indeed, the Einstein's relation was derived in the case of a hard sphere suspension in diluted regime. While the spherulites may behave like a dilute suspension during the early stages of crystallization (i.e., $t<<t_g$), the spherulites grow and interact to form network for $t \geq t_g$, with numerous relaxation modes~\cite{Song2023}.

Finally, we surprisingly observe that the elasticity of the final gels (Fig.~\ref{fig:quench}(f-Inset)) is not significantly affected by the quench temperature. One could have indeed expected that gels formed at lower temperatures would exhibit greater elasticity due to the formation of denser and smaller spherulites~\cite{shi2009} and to an increase in the concentration of the gelator incorporated in the gel network.
We  therefore suspect that the elasticity of the gel stems from both the spherulite elasticity and the way they connect to one another. While the spherulite elasticity certainly increases with lower quench temperatures as they become denser, we hypothesize that their connections become weaker. In other words, the spherulites densification occurs at the expense of the connectivity between spherulites leading to an elastisticity which does not vary significantly with the quench temperature.

\section{Conclusion}
In conclusion, this study sheds light on the formation and characteristics of viologen-based supramolecular crystal gels formed by self-assembly of low molecular weight gelators into interconnected crystalline fibers, resulting in a three-dimensional soft solid network. 
Our in-depth analyzes of the gelation kinetics triggered by thermal quenching reveal the mechanism at play to form the 3D network.
We first confirm that the viologen-based gelator $Chol_2V^{2+}$ crystallizes into nanometer-radius hollow tubes, which further assemble into micro to millimetric spherulites. Then, we demonstrate the spherulite growth is consistent with the Avrami theory. Local characterization of the growth kinetics  using optical microscopy and USALS techniques allowed us 
to determine independently the exponents in the Avrami model (Eq.~\ref{eq:n}: $ n = n_N + d n_G$). 
It reveals that the crystallization process is driven by pre-existing nuclei ($n_N=0$), with growth controlled by interfaces ($n_G=1$) and characterized by branching resulting in the formation of fractal spherulites of fractal dimension $d$.
Finally, our results demonstrate the possibility of adjusting the gel properties by varying the quenching temperature.
Lowering the quenching temperature leads to the formation of denser and smaller spherulites. We establish that the elastic properties of the gel cannot be described by the model proposed in~\cite{Liu2001,Liu2002,Liu2014}. We speculate that the elasticity of the gel stems from both the spherulite elasticity and the way they connect to one another: the spherulites densification obtained at low quenching temperature occurs at the expense of the connectivity between spherulite. This comprehensive analysis provides valuable insights into the design and manipulation of supramolecular crystal gels.

We are now considering studying the growth of fibers between electrodes using this thermal control in combination with other stimulation: magnetic/electric field, mechanical stress etc. This could lead to materials with controlled organization and orientation of the fibers opening up very promising perspective for the study and modulation of the electronic conductivity in this type of supramolecular assemblies.


\section{Appendix}

\subsection{Experimental conditions used to determine $C_{eq}$}
\label{appendix:ceq}

In order to measure the supersaturation concentration $C_{eq}$ in Fig.~\ref{fig:super}(a), samples of same initial composition ([$VBr_2$] = 1.3 mM, [$LiTFSI$] = 100 mM, [$TBAI$] = 50 mM in pentanol) were quenched at different temperatures (20°C, 30°C, 40°C, 45°C, 50°C and 60°C) then filtered (Rotilabo-syringe filter, PTFE filter, 0.20 µm pore size) to recuperate the gelator that did not participate to gel network. All samples were prepared in Eppendorf 2mL tube and were homogenized at 80°C and 1500 rpm in the Eppendorf Thermomix C (this device will be called Thermomix1), the obtained hot solutions were then maintained at 80°C. A second Eppendorf Thermomix device was set at to the quench temperature $T$ (this device will be called Themomix2), Teflon filters were at all time stored in Themomix2. In order to perform the filtration at thermal equilibrium at $T$ we used the following protocole. A plastic syringe was left for 10 minutes at 80°C in Thermomix1, then the syringe was used to take 0.9mL of hot solution. The full syringe was then quickly placed in Thermomix2, set at a temperature $T$, the syringe was maintained at this temperature for a time $t_{therm}$ allowing thermal equilibrium to be reached. Once thermal equilibrium was reached the sample was filtered on the thermalized Teflon filter, the filtrate was recovered and diluted (from a factor between 1 and 10). An absorbance spectrum $A_{filtered}$ was measured from the diluted filtrate, the measured absorbance spectra was then multiplied by the applied dilution factor in order to obtain the corrected absorbance data in Fig.~\ref{fig:super}(a). One of the samples prepared at 80°C was not filtered and simply diluted by a factor 10 in order to be used as a reference, the obtained corrected absorbance is noted $A_{ref}$. The value of $C_{eq}(T)$ corresponding to the solubility of $Chol_2V^{2+}$ was then calculated from the corrected absorbance values at 265nm (corresponding to $\pi-\pi^*$ absorption band of the viologen) using the relation: $C_{eq}=(A_{filtered}(T))/A_{ref}   \times 1.3$~mM. The time needed to reach equilibrium was estimated through the rheological measurements of growth kinetics reported in Fig.~\ref{fig:rheoKinetics}(a) and is noted $t_{min}$. The time $t_{therm}$  were chosen accordingly to the $t_{min}$ values such that $t_{therm}>t_{min}$ for all temperatures. The sample corresponding to the 60°C data was left overnight (22h) at 60°C before filtration. All the data from these experiments are available in Table Tab.~\ref{tab:1}.

\begin{table*}[t]
\fontsize{8}{8}\selectfont
  \centering
    \begin{tabular}{cccccccc}
        \hline
        T  & Sample  & t\textsubscript{min} & t\textsubscript{therm} & Dilution  & A\textsubscript{265nm}  & A\textsubscript{265nm}  & C\textsubscript{eq}  \\
        (°C) &  processing & (min) & (min)  &   factor &  measured   &   corrected &  (mM) \\\hline
        20& centrifugation & 5 & 10 & 1 & 0.515 & 0.515 & 0.19 \\
        20& filtration & 5 & 10 & 1 & 0.520 & 0.520 & 0.19 \\
        30& filtration & 5  & 10 & 2 & 0.48 & 0.96 & 0.35 \\
        40 & filtration & 8  & 10  & 2 & 0.62 & 1.24 & 0.45 \\
        45 & filtration & 34  & 45 & 4 & 0.44 & 1.76 & 0.64 \\
        50 & filtration & 75  & 140 & 4 & 0.52 & 2.08 & 0.75 \\
        60 & filtration & / & 1320 & 10 & 0.340 & 3.40 & 1.22 \\
        80& none & / & 60 & 10 & 0.36 & 3.60 & 1.3 \\
        \hline
    \end{tabular}
  \caption{Data corresponding to $C_{eq}$ measurements}
  \label{tab:1}
\end{table*}

\subsection{Radius of the spherulites over time}
\label{appendix:radius}
\begin{figure}[ht]
    \includegraphics[scale=0.5, clip=true, trim=0mm 0mm 0mm 0mm]{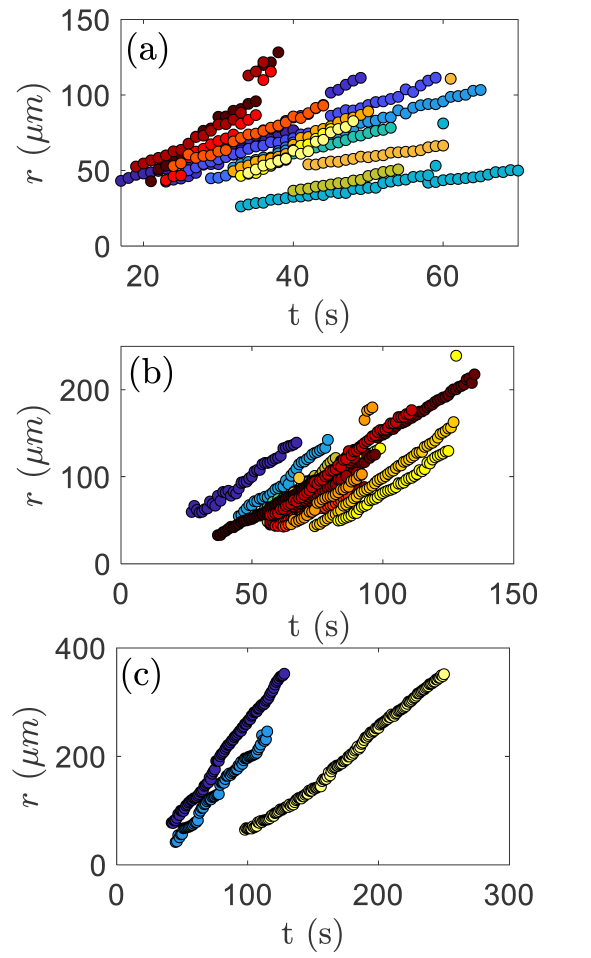}
    \centering
    \caption{Evolution of the radius of the spherulites obtained using bright field microscopy for samples at (a) $T=35$, (b) $40$ and (c) $45~^{\circ}$C.}
    \label{fig:SI_micro}
\end{figure}

Average quantities in Fig.\ref{fig:microscopy} are calculated from the measurements shown in Fig.\ref{fig:SI_micro}, where we illustrate the evolution of the radius of the spherulites as a function of time. These measurements were obtained using bright field microscopy and determined through automatic image processing.

\subsection{Absorption spectra of the viologen Iodine $VI_2$ used to determine the Beer Lambert law.}
\label{appendix:Beer}
\begin{figure}[ht]
    \includegraphics[scale=.5, clip=true, trim=0mm 0mm 0mm 0mm]{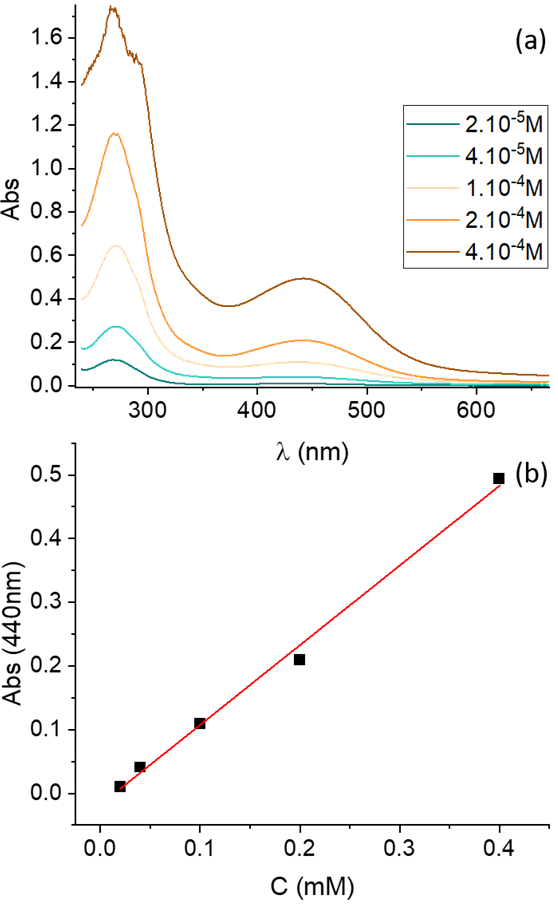}
    \centering
    \caption{a) Absorption spectra obtained at 20°C from $ Chol_2VI_2$ solutions and gels in pentanol (2mm quartz cuvette). Samples were prepared after applying a heating cooling cycle to $ Chol_2VI_2$ suspensions in 1-Pentanol. b) Evolution of the Absorbance at 440~nm with the concentration in $ Chol_2VI_2$. The continuous red line to the best linear fit.}
    \label{figS:beer}
\end{figure}

In order to characterize the formation of ${Chol_2V^{2+}/I^-}$ CT complexes in the absence of any other anionic species the viologen salt $ Chol_2VI_2$ was prepared using the following protocol. Compound $ Chol_2VBr_2$ (0.410 g, 0.32 mmol), $LiTFSI$ (3.494 g, 6.4 mmol) and 400 mL technical $EtOH$ were introduced in a 1L round bottom flask and refluxed overnight under stirring. ~80\% of the solvent was evaporated under reduced pressure, 200mL water were added and the mix was left in the fridge overnight. The mix was filtered on and washed with 3x40mL water. The white solid was then dried under vacuum, 420mg of colorless solid was recovered. This solid was dissolved in 20 mL acetone, Tetrabuthylammonium iodide (820 mg, 2.2mmol) were added and the mix was stirred at room temperature for 1h. The mix was filtered and washed with acetone (3x20mL). $Chol_2VI_2$ was obtained as an orange solid in 73\% yield. The evolution of the UV-Vis absorption spectrum of $Chol_2VI_2$ is displayed in Fig.~\ref{figS:beer}(a). In Fig.~\ref{figS:beer}(b), we observe that the absorption at $\lambda=440$~nm increase linearly with $Chol_2VI_2$ concentration leading to an adsorption coefficient  is $\epsilon_{\lambda=450nm}^{path=1 cm}=6.3 \pm 0.3 \times 10^3$ mol/L/cm. 

\subsection{Absorption peak at $\lambda=450$~nm}
\label{appendix:abs}
\begin{figure}[ht]
    \includegraphics[scale=0.9, clip=true, trim=0mm 0mm 0mm 0mm]{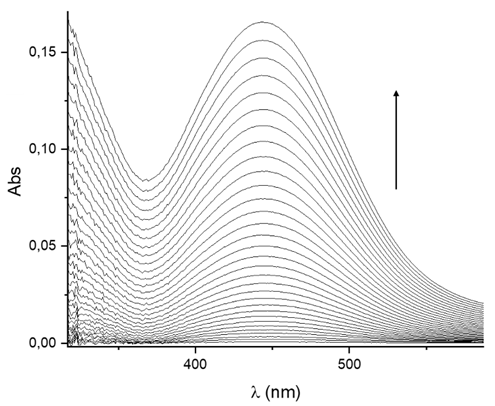}
    \centering
    \caption{Evolution of sample absorption during the first minute following nucleation at 35°C. The absorption signal at $t = t_N$ was used as a baseline and subtracted from the other spectra. The experiment corresponds to the data displayed  Fig.~\ref{fig:UvKinetics}(a).
    }
    \label{figS:abspeak}
\end{figure}

Fig.~\ref{figS:abspeak} represents the results of a slight modification of the data analysis of the absorption spectrum plotted in Fig.~\ref{fig:UvKinetics}(a). In Fig.~\ref{figS:abspeak}, we have indeed subtracted to the absorption $A(t)$ the absorption signal acquired at $t = t_N$ to highlight that the evolution of the $[Chol_2V_2^+/I^- ]$ CT band observed upon gelation corresponds to the growth of a unique absorption signal centered at $\lambda = 450$ nm.

\section*{Author contributions}
JB, VA, FP and TG carried out the experiments. JB and VA analysed the data. JB and TG wrote the article. DF, CB and TG designed and managed the project. All authors participated to the discussion of the data and the editing of the text.
\section*{Conflicts of interest}
There are no conflicts to declare.
\section*{Acknowledgment}
This work was supported by the Région Auvergne-Rhône-Alpes ``Pack Ambition Recherche", the LABEX iMUST (ANR-10-LABX-0064) of Université de Lyon, within the program "Investissements d'Avenir" (ANR-11-IDEX-0007), the ANR grants (ANR-18-CE06-0013 and ANR-21-CE06-0020-01) and by the "Mission pour les initiatives transverses et interdisciplinaires" MITI CNRS. This work benefited from meetings within the French working group GDR CNRS 2019 ``Solliciter LA Matière Molle" (SLAMM) and Block Allocation Group (BAG) at the beamline SWING with Javier Perez at Soleil obtained via the GDR SLAMM. We thank the Centre Technologique des Microstructures (CT$\mu$), Université Claude Bernard Lyon 1, Villeurbanne, France, for the CryoTEM measurements. We acknowledge the Ecole Normale Supérieure de Lyon
(ENSL) and the Centre National de la Recherche Scientifique
(CNRS) for financial, logistical, and administrative supports. V. A. thanks ENSL for the CDSN PhD Grant.
\section*{Data availability}
Data and/or code will be made available on request by contacting the authors.


\begin{thebibliography}{10}

\bibitem{ajayaghosh2008}
A.~Ajayaghosh, V.~K. Praveen, and C.~Vijayakumar.
\newblock Organogels as scaffolds for excitation energy transfer and light harvesting.
\newblock {\em Chemical Society Reviews}, 37(1):109--122, 2008.

\bibitem{andrieux2023}
V.~Andrieux, T.~Gibaud, J.~Bauland, T.~Divoux, S.~Manneville, S.~Guy, A.~Bensalah-Ledoux, L.~Guy, F.~Chevallier, D.~Frath, and C.~Bucher.
\newblock Chiral and conductive viologen-based supramolecular gels exhibiting tunable charge-transfer properties.
\newblock {\em J. Mater. Chem. C}, 11:12764--12775, 2023.

\bibitem{avrami1941}
M.~Avrami.
\newblock Granulation, phase change, and microstructure kinetics of phase change. iii.
\newblock {\em The Journal of chemical physics}, 9(2):177--184, 1941.

\bibitem{carretti2010}
E.~Carretti, M.~Bonini, L.~Dei, B.~H. Berrie, L.~V. Angelova, P.~Baglioni, and R.~G. Weiss.
\newblock New frontiers in materials science for art conservation: responsive gels and beyond.
\newblock {\em Accounts of chemical research}, 43(6):751--760, 2010.

\bibitem{chen2021}
H.~Chen, V.~Brasiliense, J.~Mo, L.~Zhang, Y.~Jiao, Z.~Chen, L.~O. Jones, G.~He, Q.-H. Guo, X.-Y. Chen, et~al.
\newblock Single-molecule charge transport through positively charged electrostatic anchors.
\newblock {\em Journal of the American Chemical Society}, 143(7):2886--2895, 2021.

\bibitem{Chowdhury2021}
S.~Chowdhury, Q.~Reynard-Feytis, C.~Roizard, D.~Frath, F.~Chevallier, C.~Bucher, and T.~Gibaud.
\newblock {Light-Controlled Aggregation and Gelation of Viologen-Based Coordination Polymers}.
\newblock {\em Journal of Physical Chemistry B}, 125(43):12063--12071, 2021.

\bibitem{Datta2016}
S.~Datta and S.~Bhattacharya.
\newblock Carbon-nanotube-mediated electrochemical transition in a redox-active supramolecular hydrogel derived from viologen and an l-alanine-based amphiphile.
\newblock {\em Chemistry – A European Journal}, 22(22):7524--7532, 2016.

\bibitem{Datta2017}
S.~Datta, N.~Dey, and S.~Bhattacharya.
\newblock Electrochemical probing of hydrogelation induced by the self-assembly of a donor–acceptor complex comprising pyranine and viologen.
\newblock {\em Chem. Commun.}, 53:2371--2374, 2017.

\bibitem{de2009}
T.~F. De~Greef, M.~M. Smulders, M.~Wolffs, A.~P. Schenning, R.~P. Sijbesma, and E.~W. Meijer.
\newblock Supramolecular polymerization.
\newblock {\em Chemical Reviews}, 109(11):5687--5754, 2009.

\bibitem{Dhiman2020}
S.~Dhiman, K.~Jalani, and S.~J. George.
\newblock Redox-mediated, transient supramolecular charge-transfer gel and ink.
\newblock {\em ACS Applied Materials \& Interfaces}, 12(5):5259--5264, 2020.

\bibitem{Dzhardimalieva2020}
G.~I. Dzhardimalieva, B.~C. Yadav, S.~Singh, and I.~E. Uflyand.
\newblock Self-healing and shape memory metallopolymers: state-of-the-art and future perspectives.
\newblock {\em Dalton Trans.}, 49:3042--3087, 2020.

\bibitem{Fang2018}
W.~Fang, Y.~Zhang, J.~Wu, C.~Liu, H.~Zhu, and T.~Tu.
\newblock Recent advances in supramolecular gels and catalysis.
\newblock {\em Chemistry – An Asian Journal}, 13(7):712--729, 2018.

\bibitem{gila2021}
C.~Gila-Vilchez, M.~C. Ma{\~n}as-Torres, J.~A. Gonz{\'a}lez-Vera, F.~Franco-Montalban, J.~A. Tamayo, F.~Conejero-Lara, J.~M. Cuerva, M.~T. Lopez-Lopez, A.~Orte, and L.~{\'A}. de~Cienfuegos.
\newblock Insights into the co-assemblies formed by different aromatic short-peptide amphiphiles.
\newblock {\em Polymer Chemistry}, 12(47):6832--6845, 2021.

\bibitem{han2020}
Y.~Han, C.~Nickle, Z.~Zhang, H.~P. Astier, T.~J. Duffin, D.~Qi, Z.~Wang, E.~Del~Barco, D.~Thompson, and C.~A. Nijhuis.
\newblock Electric-field-driven dual-functional molecular switches in tunnel junctions.
\newblock {\em Nature materials}, 19(8):843--848, 2020.

\bibitem{heijna2007}
M.~C. Heijna, M.~J. Theelen, W.~J. van Enckevort, and E.~Vlieg.
\newblock Spherulitic growth of hen egg-white lysozyme crystals.
\newblock {\em The Journal of Physical Chemistry B}, 111(7):1567--1573, 2007.

\bibitem{jones2016}
C.~D. Jones and J.~W. Steed.
\newblock {Gels with sense: Supramolecular materials that respond to heat, light and sound}, 2016.

\bibitem{Kahlfuss2018}
C.~Kahlfuss, T.~Gibaud, S.~Denis-Quanquin, S.~Chowdhury, G.~Royal, F.~Chevallier, E.~Saint-Aman, and C.~Bucher.
\newblock {Redox-Induced Molecular Metamorphism Promoting a Sol/Gel Phase Transition in a Viologen-Based Coordination Polymer}.
\newblock {\em Chemistry - A European Journal}, 24(49):13009--13019, 2018.

\bibitem{kartha2012}
K.~K. Kartha, S.~S. Babu, S.~Srinivasan, and A.~Ajayaghosh.
\newblock Attogram sensing of trinitrotoluene with a self-assembled molecular gelator.
\newblock {\em Journal of the american chemical society}, 134(10):4834--4841, 2012.

\bibitem{kathiresan2021}
M.~Kathiresan, B.~Ambrose, N.~Angulakshmi, D.~E. Mathew, D.~Sujatha, and A.~M. Stephan.
\newblock Viologens: a versatile organic molecule for energy storage applications.
\newblock {\em Journal of Materials Chemistry A}, 9(48):27215--27233, 2021.

\bibitem{lai2016}
Z.~Lai, Y.~Chen, C.~Tan, X.~Zhang, and H.~Zhang.
\newblock Self-assembly of two-dimensional nanosheets into one-dimensional nanostructures.
\newblock {\em Chem}, 1(1):59--77, 2016.

\bibitem{lakshminarayanan2021}
V.~Lakshminarayanan, C.~Chockalingam, E.~Mendes, and J.~H. van Esch.
\newblock Gelation kinetics-structure analysis of ph-triggered low molecular weight hydrogelators.
\newblock {\em ChemPhysChem}, 22(21):2256--2261, 2021.

\bibitem{lam2010}
R.~Lam, L.~Quaroni, T.~Pedersen, and M.~A. Rogers.
\newblock A molecular insight into the nature of crystallographic mismatches in self-assembled fibrillar networks under non-isothermal crystallization conditions.
\newblock {\em Soft Matter}, 6(2):404--408, 2010.

\bibitem{landman2018}
J.~Landman, S.~Ouhajji, S.~Pr{\'e}vost, T.~Narayanan, J.~Groenewold, A.~P. Philipse, W.~K. Kegel, and A.~V. Petukhov.
\newblock Inward growth by nucleation: Multiscale self-assembly of ordered membranes.
\newblock {\em Science advances}, 4(6):eaat1817, 2018.

\bibitem{lescanne2003}
M.~Lescanne, A.~Colin, O.~Mondain-Monval, F.~Fages, and J.-L. Pozzo.
\newblock Structural aspects of the gelation process observed with low molecular mass organogelators.
\newblock {\em Langmuir}, 19(6):2013--2020, 2003.

\bibitem{li2021}
J.~Li, S.~Pudar, H.~Yu, S.~Li, J.~S. Moore, J.~Rodr{\'\i}guez-L{\'o}pez, N.~E. Jackson, and C.~M. Schroeder.
\newblock Reversible switching of molecular conductance in viologens is controlled by the electrochemical environment.
\newblock {\em The Journal of Physical Chemistry C}, 125(40):21862--21872, 2021.

\bibitem{Li2010}
J.~L. Li and X.~Y. Liu.
\newblock {Architecture of supramolecular soft functional materials: From understanding to micro-/nanoscale engineering}.
\newblock {\em Advanced Functional Materials}, 20(19):3196--3216, 2010.

\bibitem{Liu2001}
X.~Y. Liu and P.~D. Sawant.
\newblock {Formation kinetics of fractal nanofiber networks in organogels}.
\newblock {\em Applied Physics Letters}, 79(21):3518--3520, 2001.

\bibitem{Liu2002}
X.~Y. Liu and P.~D. Sawant.
\newblock {Determination of the fractal characteristic of nanofiber-network formation in supramolecular materials}.
\newblock {\em ChemPhysChem}, 3(4):374--377, 2002.

\bibitem{liu_sawant2002}
X.~Y. Liu, P.~D. Sawant, W.~B. Tan, I.~Noor, C.~Pramesti, and B.~Chen.
\newblock Creating new supramolecular materials by architecture of three-dimensional nanocrystal fiber networks.
\newblock {\em Journal of the American Chemical Society}, 124(50):15055--15063, 2002.

\bibitem{Liu2014}
Y.~Liu, R.~Y. Wang, J.~L. Li, B.~Yuan, M.~Han, P.~Wang, and X.~Y. Liu.
\newblock {Identify kinetic features of fibers growing, branching, and bundling in microstructure engineering of crystalline fiber network}.
\newblock {\em CrystEngComm}, 16(24):5402--5408, 2014.

\bibitem{livsey1987}
I.~Livsey.
\newblock Neutron scattering from concentric cylinders. intraparticle interference function and radius of gyration.
\newblock {\em Journal of the Chemical Society, Faraday Transactions 2: Molecular and Chemical Physics}, 83(8):1445--1452, 1987.

\bibitem{ma2008}
Z.~Ma, G.~Zhang, X.~Zhai, L.~Jin, X.~Tang, M.~Yang, P.~Zheng, and W.~Wang.
\newblock Fractal crystal growth of poly (ethylene oxide) crystals from its amorphous monolayers.
\newblock {\em Polymer}, 49(6):1629--1634, 2008.

\bibitem{mao2016}
B.~Mao, T.~Divoux, and P.~Snabre.
\newblock Normal force controlled rheology applied to agar gelation.
\newblock {\em Journal of Rheology}, 60(3):473--489, 2016.

\bibitem{mareau2005}
V.~H. Mareau and R.~E. Prud'Homme.
\newblock In-situ hot stage atomic force microscopy study of poly ($\varepsilon$-caprolactone) crystal growth in ultrathin films.
\newblock {\em Macromolecules}, 38(2):398--408, 2005.

\bibitem{monk1999}
P.~M. Monk, N.~M. Hodgkinson, and R.~D. Partridge.
\newblock The colours of charge-transfer complexes of methyl viologen: effects of donor, ionic strength and solvent.
\newblock {\em Dyes and pigments}, 43(3):241--251, 1999.

\bibitem{murata1994}
K.~Murata, M.~Aoki, T.~Suzuki, T.~Harada, H.~Kawabata, T.~Komori, F.~Ohseto, K.~Ueda, and S.~Shinkai.
\newblock Thermal and light control of the sol-gel phase transition in cholesterol-based organic gels. novel helical aggregation modes as detected by circular dichroism and electron microscopic observation.
\newblock {\em Journal of the American Chemical Society}, 116(15):6664--6676, 1994.

\bibitem{nasr2021}
P.~Nasr, H.~Leung, F.-I. Auzanneau, and M.~A. Rogers.
\newblock Supramolecular fractal growth of self-assembled fibrillar networks.
\newblock {\em Gels}, 7(2):46, 2021.

\bibitem{Newbloom2012}
G.~M. Newbloom, K.~M. Weigandt, and D.~C. Pozzo.
\newblock {Electrical, mechanical, and structural characterization of self-assembly in poly(3-hexylthiophene) organogel networks}.
\newblock {\em Macromolecules}, 45(8):3452--3462, 2012.

\bibitem{nguyen2018}
Q.~V. Nguyen, P.~Martin, D.~Frath, M.~L. Della~Rocca, F.~Lafolet, S.~Bellinck, P.~Lafarge, and J.-C. Lacroix.
\newblock Highly efficient long-range electron transport in a viologen-based molecular junction.
\newblock {\em Journal of the American Chemical Society}, 140(32):10131--10134, 2018.

\bibitem{pernetti2007}
M.~Pernetti, K.~van Malssen, D.~Kalnin, and E.~Fl{\"o}ter.
\newblock Structuring edible oil with lecithin and sorbitan tri-stearate.
\newblock {\em Food Hydrocolloids}, 21(5-6):855--861, 2007.

\bibitem{Piau1999}
J.-M. Piau, M.~Dorget, J.-F. Palierne, and A.~Pouchelon.
\newblock {Shear elasticity and yield stress of silica–silicone physical gels: Fractal approach}.
\newblock {\em Journal of Rheology}, 43(2):305--314, mar 1999.

\bibitem{Rao2010}
K.~V. Rao, K.~Jayaramulu, T.~K. Maji, and S.~J. George.
\newblock Supramolecular hydrogels and high-aspect-ratio nanofibers through charge-transfer-induced alternate coassembly.
\newblock {\em Angewandte Chemie International Edition}, 49(25):4218--4222.

\bibitem{Roizard2022}
C.~Roizard, V.~Andrieux, S.~A. Shehimy, S.~Chowdhury, Q.~Reynard-Feytis, C.~Kahlfuss, E.~Saint-Aman, F.~Chevallier, C.~Bucher, T.~Gibaud, and D.~Frath.
\newblock Photoredox processes in the aggregation and gelation of electron-responsive supramolecular polymers based on viologen.
\newblock {\em ECS Advances}, 1(2):020502, 2022.

\bibitem{sarkar2020}
A.~Sarkar, R.~Sasmal, C.~Empereur-Mot, D.~Bochicchio, S.~V. Kompella, K.~Sharma, S.~Dhiman, B.~Sundaram, S.~S. Agasti, G.~M. Pavan, et~al.
\newblock Self-sorted, random, and block supramolecular copolymers via sequence controlled, multicomponent self-assembly.
\newblock {\em Journal of the American Chemical Society}, 142(16):7606--7617, 2020.

\bibitem{saunders2019}
L.~Saunders and P.~X. Ma.
\newblock Self-healing supramolecular hydrogels for tissue engineering applications.
\newblock {\em Macromolecular bioscience}, 19(1):1800313, 2019.

\bibitem{shi2009}
J.~H. Shi, X.~Y. Liu, J.~L. Li, C.~S. Strom, and H.~Y. Xu.
\newblock Spherulitic networks: from structure to rheological property.
\newblock {\em The journal of physical chemistry b}, 113(14):4549--4554, 2009.

\bibitem{shirzad2023}
K.~Shirzad and C.~Viney.
\newblock A critical review on applications of the avrami equation beyond materials science.
\newblock {\em Journal of the Royal Society Interface}, 20(203):20230242, 2023.

\bibitem{shtukenberg2012}
A.~G. Shtukenberg, Y.~O. Punin, E.~Gunn, and B.~Kahr.
\newblock Spherulites.
\newblock {\em Chemical reviews}, 112(3):1805--1838, 2012.

\bibitem{Song2023}
J.~Song, N.~Holten-Andersen, and G.~H. McKinley.
\newblock {Non-Maxwellian viscoelastic stress relaxations in soft matter}.
\newblock {\em Soft Matter}, 19(41):7885--7906, oct 2023.

\bibitem{striepe2017}
L.~Striepe and T.~Baumgartner.
\newblock Viologens and their application as functional materials.
\newblock {\em Chemistry--A European Journal}, 23(67):16924--16940, 2017.

\bibitem{Suzuki2001}
M.~Suzuki, C.~C. Waraksa, H.~Nakayama, K.~Hanabusa, M.~Kimura, and H.~Shirai.
\newblock Supramolecular assemblies formed by new l-lysine derivatives of viologens.
\newblock {\em Chem. Commun.}, pages 2012--2013, 2001.

\bibitem{Sztucki2007}
M.~Sztucki and T.~Narayanan.
\newblock {Development of an ultra-small-angle X-ray scattering instrument for probing the microstructure and the dynamics of soft matter}.
\newblock {\em Journal of Applied Crystallography}, 40(s1):s459--s462, Apr 2007.

\bibitem{taguchi2001}
K.~Taguchi, H.~Miyaji, K.~Izumi, A.~Hoshino, Y.~Miyamoto, and R.~Kokawa.
\newblock Growth shape of isotactic polystyrene crystals in thin films.
\newblock {\em Polymer}, 42(17):7443--7447, 2001.

\bibitem{thureau2021}
A.~Thureau, P.~Roblin, and J.~P{\'e}rez.
\newblock Biosaxs on the swing beamline at synchrotron soleil.
\newblock {\em Journal of Applied Crystallography}, 54(6):1698--1710, 2021.

\bibitem{van2003}
K.~J. Van~Bommel, A.~Friggeri, and S.~Shinkai.
\newblock Organic templates for the generation of inorganic materials.
\newblock {\em Angewandte Chemie International Edition}, 42(9):980--999, 2003.

\bibitem{wang2006}
R.~Wang, X.-Y. Liu, J.~Xiong, and J.~Li.
\newblock Real-time observation of fiber network formation in molecular organogel: supersaturation-dependent microstructure and its related rheological property.
\newblock {\em The journal of physical chemistry b}, 110(14):7275--7280, 2006.

\bibitem{webber2017}
M.~J. Webber and R.~Langer.
\newblock Drug delivery by supramolecular design.
\newblock {\em Chemical Society Reviews}, 46(21):6600--6620, 2017.

\bibitem{gels2006}
R.~G. Weiss and P.~Terech.
\newblock Molecular gels.
\newblock {\em Materials with Self-Assembled Fibrillar Networks}, 2006.

\bibitem{Xue2004}
P.~Xue, R.~Lu, D.~Li, M.~Jin, C.~Bao, Y.~Zhao, and Z.~Wang.
\newblock {Rearrangement of the aggregation of the gelator during sol-gel transcription of a dimeric cholesterol-based viologen derivative into fibrous silica}.
\newblock {\em Chem. Mater.}, 16(19):3702--3707, sep 2004.

\bibitem{yoshikawa2005}
H.~Yoshikawa and S.-i. Nishikiori.
\newblock Crystal structures and spectroscopic properties of polycyano--polycadmate host clathrates including a ct complex guest of methylviologen dication and aromatic donor.
\newblock {\em Dalton Transactions}, (18):3056--3064, 2005.

\bibitem{yu2015}
R.~Yu, N.~Lin, W.~Yu, and X.~Y. Liu.
\newblock {Crystal networks in supramolecular gels: Formation kinetics and mesoscopic engineering principles}.
\newblock {\em CrystEngComm}, 17(42):7986--8010, 2015.

\bibitem{yuan2011}
B.~Yuan, X.-Y. Liu, J.-L. Li, and H.-Y. Xu.
\newblock Volume confinement induced microstructural transitions and property enhancements of supramolecular soft materials.
\newblock {\em Soft Matter}, 7(5):1708--1713, 2011.

\bibitem{Yuan2016}
T.~Yuan, M.~Vazquez, A.~N. Goldner, Y.~Xu, R.~Contrucci, M.~A. Firestone, M.~A. Olson, and L.~Fang.
\newblock Versatile thermochromic supramolecular materials based on competing charge transfer interactions.
\newblock {\em Advanced Functional Materials}, 26(47):8604--8612.

\bibitem{Yuan2017}
T.~Yuan, Y.~Xu, C.~Zhu, Z.~Jiang, H.-J. Sue, L.~Fang, and M.~A. Olson.
\newblock Tunable thermochromism of multifunctional charge-transfer-based supramolecular materials assembled in water.
\newblock {\em Chemistry of Materials}, 29(23):9937--9945, 2017.

\bibitem{zhu2007}
D.-S. Zhu, Y.-X. Liu, E.-Q. Chen, M.~Li, C.~Chen, Y.-H. Sun, A.-C. Shi, R.~M. Van~Horn, and S.~Z. Cheng.
\newblock Crystal growth mechanism changes in pseudo-dewetted poly (ethylene oxide) thin layers.
\newblock {\em Macromolecules}, 40(5):1570--1578, 2007.

\end{thebibliography}

\end{document}